\documentclass[aps,prb,twocolumn,showpacs,superscriptaddress]{revtex4}
\usepackage{amssymb,amsmath}
\usepackage{graphicx}
\usepackage{multirow}
\usepackage{hyperref}
\usepackage{appendix}
\usepackage{mathtools}
\usepackage{amsmath}
\usepackage{siunitx}
\graphicspath{{./}}

\begin{document}

\title{First-principles Study of the Luminescence of Eu$^{2+}$-doped Phosphors}

\author{Yongchao Jia}
\email[]{yongchao.jia@uclouvain.be}
\affiliation{European Theoretical Spectroscopy Facility, Institute of Condensed Matter and Nanosciences, Universit\'{e} catholique de Louvain, Chemin des \'{e}toiles 8, bte L07.03.01, B-1348 Louvain-la-Neuve, Belgium}
\author{Anna  Miglio}
\affiliation{European Theoretical Spectroscopy Facility, Institute of Condensed Matter and Nanosciences, Universit\'{e} catholique de Louvain, Chemin des \'{e}toiles 8, bte L07.03.01, B-1348 Louvain-la-Neuve, Belgium}
\author{Samuel Ponc\'{e}}
\affiliation{Department of Materials, University of Oxford, Parks Road, Oxford, OX1 3PH, UK}
\author{Masayoshi Mikami}
\affiliation{Functional Materials Design Laboratory, Yokohama R$\&$D Center, Mitsubishi Chemical
Corporation, 1000,
Kamoshida-cho Aoba-ku, Yokohama, 227-8502, Japan}
\author{Xavier Gonze}
\affiliation{European Theoretical Spectroscopy Facility, Institute of Condensed Matter and Nanosciences, Universit\'{e} catholique de Louvain, Chemin des \'{e}toiles 8, bte L07.03.01, B-1348 Louvain-la-Neuve, Belgium}

\date{\today}

\begin{abstract}
The luminescence of fifteen representative Eu$^{2+}$-doped phosphors used for white-LED and scintillation applications is studied through a Constrained Density Functional Theory. Transition energies and Stokes shift are deduced from differences of total energies between the ground and excited states of the systems, in the absorption and emission geometries. The general applicability of such methodology is first assessed: for this representative set, the calculated absolute error with respect to experiment on absorption and emission energies is within 0.3~eV. This set of compounds covers a wide range of transition energies that extents from 1.7 to 3.5~eV. The information gained from the relaxed geometries and total energies is further used to evaluate the thermal barrier for the $4f-5d$ crossover, the full width at half-maximum of the emission spectrum and the temperature shift of the emission peak, using a one-dimensional configuration-coordinate model. The former results indicate that the $4f-5d$ crossover cannot be the dominant mechanism for the thermal quenching behavior of Eu$^{2+}$-doped phosphors and the latter results are compared to available experimental data and yield a 30$\%$ mean absolute relative error. Finally, a semi-empirical model used previously for Ce$^{3+}$-doped hosts is adapted to Eu$^{2+}$-doped hosts and gives the absorption and emission energies within 0.9~eV of experiment, underperforming compared to the first-principles calculation.
\end{abstract}


\maketitle

\section{Introduction}
\label{intro}

Eco-efficient light-emitting diodes (LEDs) are increasingly used as new-generation light sources for general white lighting, 
with blue- or UV-LED generating the highest frequency photons, and  phosphors downconverting some of these photons to lower visible frequencies.
The Nobel Prize in Physics 2014 was awarded to the blue-LED inventors in view of the physical and technological challenges they have overcome, and the impact of this achievement.  
As a key component, phosphors have an important effect on the performance of the white-LEDs, 
especially on the correlated color temperature (CCT) and color-rendering index (CRI).\cite{SLA,led-1,led-2} 
For this reason, the US Department of Energy has defined a 2020 target for the green and red-emission converters, 
which mentioned that the developed phosphors should possess a narrow emission band with high thermal stability.\cite{BSO12N2,BSO2N2,CLA,SMSN,led-3}

Accordingly, a lot of efforts have been devoted to the development of efficient rare-earth (RE) ion doped phosphors, 
especially the narrow-band green/red emission Eu$^{2+}$-doped ones. 
However, most of these efforts have relied on (semi)empirical insights. 
One typical example is the recently developed Sr[LiAl$_3$N$_4$]:Eu$^{2+}$ (denoted as SLA:Eu later), 
that might become the next-generation commercial red phosphor.\cite{SLA} 
Even though excellent optical properties such as red emission color (650~nm), small full-width at half-maximum (FWHM $\sim$50 nm), high thermal stability ($>$95\% 
relative to the quantum efficiency at 450~K) are experimentally obtained, the exact origin of these superior properties is still unknown. 
Also, two inequivalent Sr$^{2+}$ sites exist for the Eu$^{2+}$ substitutional doping in this host, 
while only a single narrow emission peak has been observed. The luminescent center in this phosphor has not been determined yet. 
To address these questions, a quantitative understanding of the optical behaviour of Eu$^{2+}$-doped phosphors, at the atomic scale, is urgently needed.

Theoretical modeling of the luminescence of RE ions in inorganic compounds dates back to the 1960s 
when the Judd-Ofelt theory was proposed to analyze 
the $4f\rightarrow4f$ transitions in RE ions by fitting parameterized Hamiltonians of $4f$ electrons.\cite{JO-1,JO-2} 
This theory has been recently extended to depict the $4f\rightarrow5d$ transition of RE ions.\cite{lf1,lf2,lf3} 
However, the complex parameter fitting procedure severely limits its usage to a small number of compounds, 
even with the aid of Ligand Field Density Functional Theory (LFDFT).\cite{werner1,werner2,werner3} 
Beside these works, several other efforts have been conducted to understand the luminescence of $4f\rightarrow5d$ neutral excitation of RE ions in inorganic compounds, 
based on ab-initio quantum chemistry finite-cluster (QCFC) method or semi-empirical analysis. \cite{q1,q2,dorenbos1,dorenbos2,dorenbos4,Wang2015} 
For example, the QCFC method has been widely used in the analysis of Ce$^{3+}$-doped materials, 
to identify the luminescent center, based on the absorption spectrum.\cite{Luis-1, Luis-2, Luis-3,q1,q2} 
For the semi-empirical analysis, the most widely used model was proposed by Dorenbos, who quantified the nephelauxetic effect 
and crystal field splitting of RE$_{5d}$ state.\cite{dorenbos1,dorenbos2,dorenbos4} 
This model provides correct general trends for the absorption properties of Ce$^{3+}$-doped phosphors. 

Despite such achievements, limitations are present in these two approaches. 
Indeed, on one hand, the QCFC method does not account for electronic and vibrational properties of the host  
which are crucial for predicting the thermal quenching behavior and FWHM of the emission peak.\cite{phonon1,phonon2,4f-5d} 
Also, this cluster method is limited to the quantitative analysis of the luminescence of Ce$^{3+}$ 
ion due to its simplest electronic configuration, $4f^05d^1$, in the excited state.\cite{werner1} 
On the other hand, for the semi-empirical method, quantitative predictions are obtained through fitted parameters for specific classes of compounds. 
These fitting procedures have only been done for fluoride, oxide and nitride-based compounds with Ce$^{3+}$ doping. 
Such analysis for the Eu$^{2+}$-doped materials is not yet available, although several quantitative relationships for the luminescence 
of Eu$^{2+}$ and Ce$^{3+}$ ions in the same inorganic host have been found.\cite{dorenbos5} 
Another problem of this semi-empirical approach is that the analysis only focuses on the absorption process, 
while the emission process and Stokes shift are not considered due to the lack of experimental data related to the relaxed excited-state crystal structure.

In our recent work, we have studied the luminescence of more than a dozen Ce$^{3+}$-doped phosphors based on first-principles calculations.\cite{jia2016,jia2017} 
In this context, we have assessed the accuracy of a theoretical methodology to obtain the transition energy and Stokes shift of Ce$^{3+}$-doped phosphors. The method is based on a Constrained Density Functional Theory (CDFT) and $\Delta$SCF analysis of the total energies. 
For the sake of brevity, we will denote this approach as $\Delta$SCF method. 
The general applicability of the $\Delta$SCF method has been investigated: 
the obtained transition energies match experimental data within 0.3 eV in general, over a range that extents from 2 to 5 eV. 
In addition, the ground and excited structural information from the $\Delta$SCF method has been used to parameterize the Dorenbos's semi-empirical model, 
and extend its predicting ability from the absorption process to the emission process.\cite{jia2017} 

In this paper, we consider similarly the $\Delta$SCF method for the analysis of the luminescence of Eu$^{2+}$ ion in inorganic materials. 
We aim to: 
(1) assess the accuracy of the $\Delta$SCF method in obtaining the absorption and emission transition energies and Stokes shifts for such Eu$^{2+}$-doped phosphors; 
(2) obtain an evaluation of the thermal energy barrier for the $4f-5d$ crossover, FWHM and the temperature shift of the $5d\rightarrow4f$ emission peak 
based on the simple one-dimensional configuration coordinate model (1D-CCD); 
(3) fit the Dorenbos' semi-empirical model for the analysis of Eu$^{2+}$-doped phosphors for both absorption and emission states, 
and compare the resulting accuracy of the $\Delta$SCF method to the Dorenbos' semi-empirical model. 

The work is thus structured as follows. In Section \ref{num_approach}, 
we first describe the $\Delta$SCF method in the case of Eu$^{2+}$-doped phosphors. 
The theoretical method to obtain the $4f-5d$ thermal barrier, FWHM and temperature shift for the emission peak is then explained. We also present the Dorenbos' semi-empirical model in the case of Eu$^{2+}$-doped phosphors. 
In Section \ref{validation and usage}, absorption, emission energies, and Stokes shifts, from the $\Delta$SCF method are first presented, and compared
with experimental data, for fifteen representative Eu$^{2+}$-doped materials. 
Then, the very same information (total energies and relaxed geometries) from the $\Delta$SCF method allows us to evaluate the energy barrier for $4f-5d$ crossover, FWHM and temperature shift of Eu$^{2+}$-doped materials within the framework of the 1D-CCD, 
and fit the proposed Dorenbos' semi-empirical model. The conclusions are given in Section \ref{conclusion}.   

\section{Methodology}
\label{num_approach}

In this section, we first present the methods and procedures that are required to compute the ground state configuration, and related numerical parameters. We then focus on the excited state. Afterwards, we introduce the configuration coordinate diagram and its by-products: optical properties, including the transition energies and Stokes shifts, approximate thermal quenching barrier for the $4f-5d$ crossover, FWHM and related temperature shift, fully from first-principles. We also extend the Dorenbos' semi-empirical model for the analysis of Eu$^{2+}$-doped phosphors.

\subsection{Ground-state calculation}
\label{gs}

The calculations have been performed within density-functional theory (DFT) using the projector-augmented wave (PAW) method as implemented in the ABINIT package.\cite{Blochl1994,Marc2008,abinit2002,abinit2008,abinit2016} 
Exchange-correlation effects were treated within the generalized gradient approximation (GGA-PBE) \cite{Perdew1996}, with the addition of a Hubbard U term for the $4f$ states of the Eu ion.\cite{dft+u} 

Such U term was crucial to obtain the experimentally well-established presence of Eu$_{4f}$ levels inside the band gap, a defining characteristics of such efficient luminescent materials.\cite{jia2016,samuel2016,Blasse} 
The U value was fixed to 7.5~eV for all fifteen representative Eu$^{2+}$-doped materials, as in our earlier study of two Eu$^{2+}$-doped barium silicate oxynitrides.\cite{samuel2016}
It was checked that deviation by up to $\pm$ 1.0~eV in the U value had little ($< 0.1$~eV) impact on the transition energies.
The same U value successfully places the Eu$_{4f}$ states in the band gap for the fifteen Eu$^{2+}$-doped materials, making the comparison with experiment free of adjustable parameters.  However,we want to point out that we also tried CaO:Eu with the same U value, and it correctly placed the Eu$_{4f}$ states in the band gap for the ground state, but not for the excited state, as seen in Section~\ref{validation}.

All the PAW atomic datasets were taken from the ABINIT website.\cite{Jollet2014} 
The calculations were based on the supercell method, in which the primitive cell of the host is repeated to form a large (non-primitive) cell, and one of the suitable cations is replaced by a Eu$^{2+}$ ion.
Structural relaxation and band structure calculations were converged to 10$^{-5}$~Hartree/Bohr (for residual forces) 
and 0.5 mHa/atom (for the tolerance on the total energy). 
In all these calculations, a kinetic-energy cutoff of 30 Ha for the plane-wave basis set was used. A larger kinetic-energy cutoff of 40 Ha has been used to test the convergence in several cases. The result indicated that the energy cutoff of 30 Ha provided a converged result within 0.1~eV for the transition energies.  
More details on the performed calculations, including the supercell size, Eu$^{2+}$ ion concentration and \textbf{k}-point grid for the fifteen representative Eu$^{2+}$-doped materials, are listed in the Appendix.

Although the spin-orbit coupling can play some role in the electronic structure of Eu compounds, it has been neglected in the present study. Indeed, for the Eu$^{2+}$ ion, the spin-orbit coupling will not yield
multiplet splitting of the ground state configuration with $4f^7$. 
By contrast, spin-orbit coupling will have a strong effect on the excited state electron configuration, $4f^65d^1$. Focusing on the $4f^6$ term, there is a multiplet splitting $^7F_J$ (J = 0-6) occurring in the absorption spectrum. However, in the present work, we only focus on the lowest energy excited state in this multiplet and do not attempt to describe the other excited states. Therefore, spin-orbit coupling was not explicitly included.

\subsection{Excited state calculation}
\label{ex}

Even though ground-state DFT+U correctly places the Eu$_{4f}$ states inside the band gap, the Eu$_{5d}$ states are not found within the band gap for most of the fifteen cases, which is opposite to the experimental results, as the Eu$^{2+}$ ion gives an efficient luminescence in all these compounds. Therefore, the optical properties of these RE-ion doped phosphors must be treated by an excited-state theory. 
The $4f\rightarrow 5d$ excitation is a neutral excitation, and the Bethe-Salpeter Equation (BSE) of Many-Body Perturbation Theory (MBPT) \cite{Onida2002} 
is considered state of the art to treat such neutral excitation. 
However, the computational load and memory needs for such approach are prohibitive for supercells of about fifty to one hundred atoms like in the present work. 
Instead of the standard BSE method, in our previous study, following the works of Canning \textit{et al}, \cite{canning2011,Chaudhry2014} we have simulated the $4f\rightarrow5d$ neutral excitation of Ce$^{3+}$ ion on the basis of the ${\Delta}$SCF approach. Although the use of the ${\Delta}$SCF approach is theoretically founded for the lowest state of each symmetry representation,\cite{Gunnarson1976} the rotational symmetry is broken here. So, like in previous studies using ${\Delta}$SCF, 
we work beyond formal justification. 
The electron-hole interaction, an essential contribution in the BSE, is mimicked by promoting the Ce$_{4f}$ electron to the Ce$_{5d}$ state: we constrain one $4f$-type band to be unoccupied, 
while occupying the lowest $5d$-type band lying higher in energy. 
Our prior study on Ce$^{3+}$-doped materials has demonstrated the ${\Delta}$SCF ability to yield quantitative predictions for the transition energies and Stokes shift, deduced from the total energy differences of the different constrained configurations.\cite{Tozera2000,Martin,phonon2}

\begin{figure}
\includegraphics[scale=0.3]{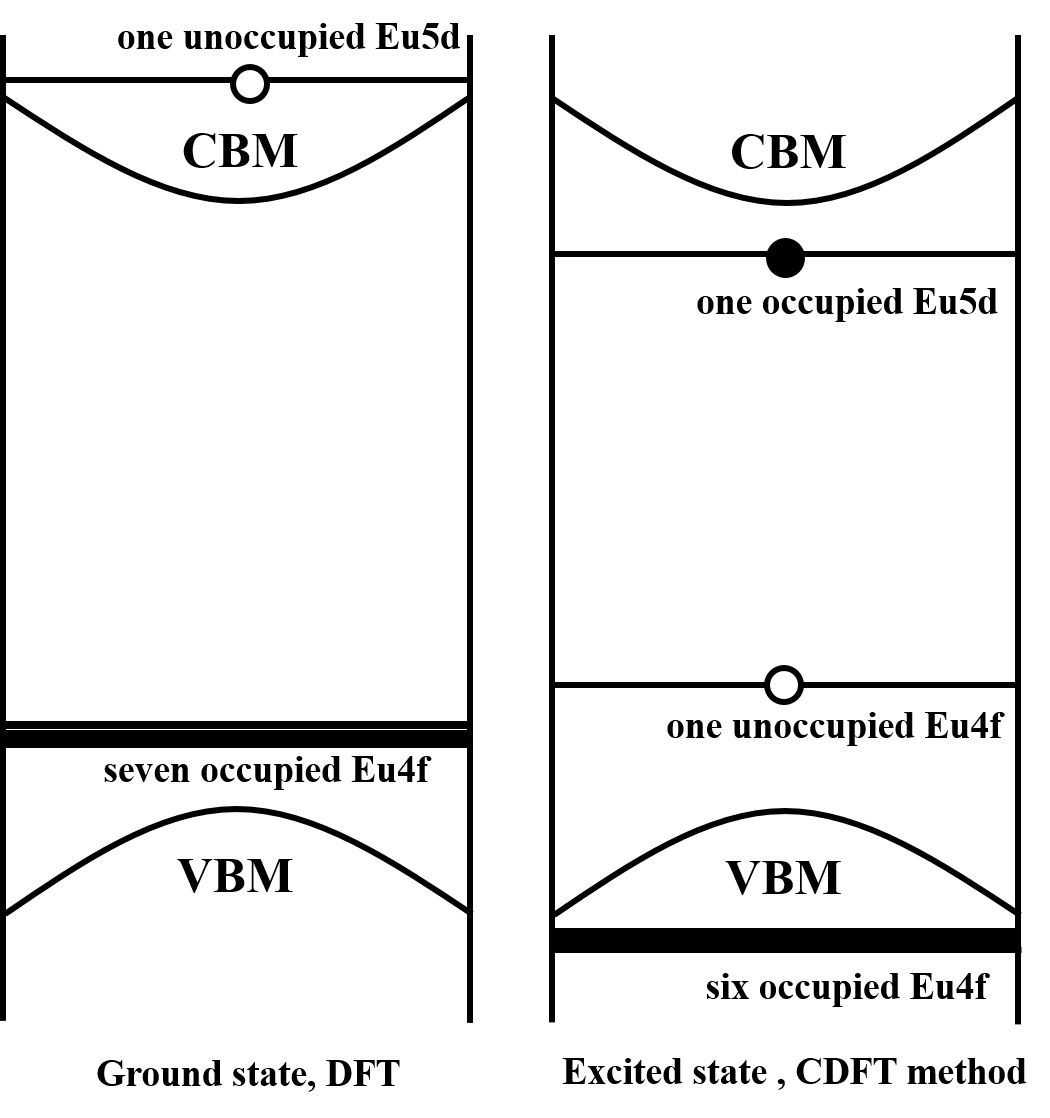}
\caption{Schematic electron occupancies for the ground and excited state calculations of Eu$^{2+}$-doped phosphors.}\label{figure_1}
\end{figure}

In this work, we apply the same $\Delta$SCF approach to Eu$^{2+}$-doped materials. 
Note that the electron configurations of the ground state of Ce$^{3+}$ and Eu$^{2+}$ ions are different. 
The Ce$^{3+}$ ion has only one single electron in the $4f$ state, while seven unpaired electrons exist in the $4f$ states of the Eu$^{2+}$ ion, corresponding to a half-filled $4f^7$ configuration.
Therefore, the usage of the ${\Delta}$SCF method is relatively straightforward in Ce$^{3+}$-doped materials, while special attention must be paid to the electron occupancy of the excited state in the study of Eu$^{2+}$-doped phosphors. 
We found that the promotion of the highest $4f$ electron of the Eu$^{2+}$ ion provides the best results for the ${\Delta}$SCF method (also the lowest energy). The alternative equal depletion of all $4f$ states by an amount of 1/7 lead to the unphysical hybridization of all the $4f$ states with the valence band of the host material.
With the promotion of the highest $4f$ electron, the $4f$-type 
bands split into one unoccupied band that stays within the gap, and six occupied bands that shift downwards, actually hybridizing inside the valence band. 
Fig.~\ref{figure_1} depicts such electron occupancies for the ground and excited-state calculations of Eu$^{2+}$-doped phosphors. We also considered CaO:Eu in addition to the list of fifteen materials, but in this case, 
all seven Eu$_{4f}$ states enter the valence band, so we cannot localize one additional hole on the Eu ion, and moreover the Eu$_{5d}$ state does not appear inside the band gap. We expect that a higher-level DFT 
approximation that is able to correct the band gap, like an hybrid functional, might be needed.

\subsection{One-dimensional configuration coordinate diagram}
\label{conf_coord}

\begin{figure}
\includegraphics[scale=0.3]{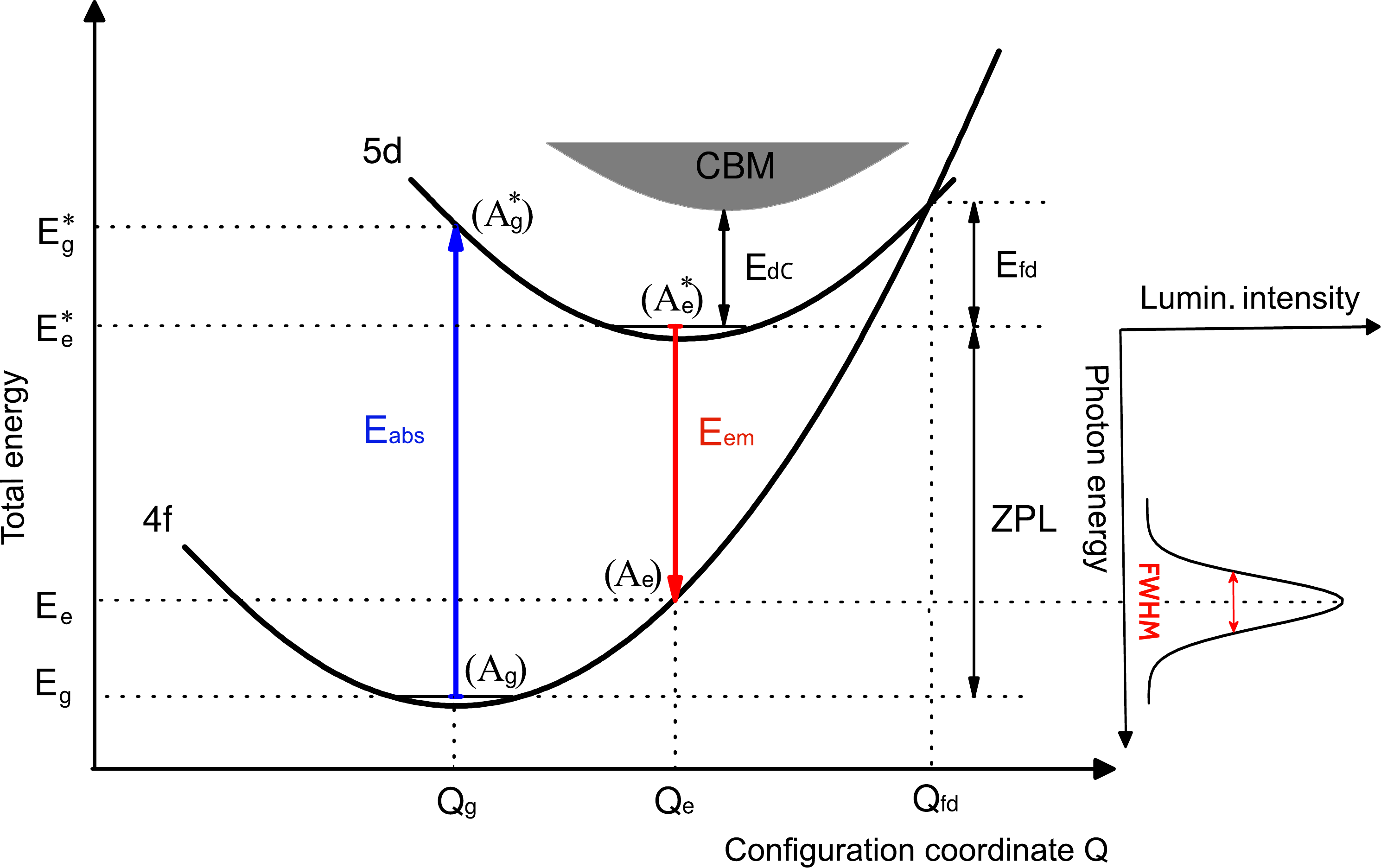}
\caption{The one-dimensional configuration coordinate diagram.}\label{figure_2}
\end{figure}

The 1D-CCD diagram shown in Fig.~\ref{figure_2}, provides a simple representation of the combined effect of electronic excitation and geometry relaxation.
It depicts the Born-Oppenheimer potential energy of a system containing Eu$^{2+}$ ion in its ground and excited state (curves $4f$ and $5d$)
respectively, as a function of one generalised configuration coordinate Q, connecting the ionic coordinates of the system for the electronic ground and excited states.

Such a one-dimensional representation ignores the full complexity of all possible collective nuclei displacements that might play a role in the detailed 
description of the luminescence process, but instead focuses on the single most relevant one.\cite{toyozawa2003optical,book1,book2} 
Q$_{\text{g}}$ and $Q_{\text{e}}$ represent the equilibrium configuration coordinates for the system with
Eu$^{2+}$ ion in its ground and excited states, respectively. The horizontal
lines inside the curves $4f$ and $5d$ denote the energy levels of the system in which the quantization of vibrational motion is taken into account. When a photon is absorbed by
the Eu$_{4f}$ electron, the Eu$^{2+}$ ion will be excited from its ground
state to the excited state, corresponding to A$_g\rightarrow$A$_g^{*}$. After the absorption, the system will be out of equilibrium due to the change in the electronic configuration of the Eu$^{2+}$ ion. 
The atomic positions are then relaxed following the forces in the electronic excited state, which is represented by the process A$_g^{*}\rightarrow$A$_e^*$ in Fig.~\ref{figure_2}.
After this lattice relaxation, the system reaches a new metastable state, at which the emission process A$_e^*\rightarrow$A$_e$ occurs. 
The cycle is completed by the lattice relaxation A$_e\rightarrow$A$_g$ in the electronic ground state. 
Based on this idea, the absorption/emission energy, Franck-Condon shifts and the Stokes shift of Eu$^{2+}$-doped phosphors can be determined semi-classically as follows.
The absorption process, with energy 
\begin{equation}
E_{\text{abs}} = E_\text{g}^\text{*} - E_\text{g},
\end{equation} 
is followed by multi-phonon emission, with Franck-Condon shift in the excited state
\begin{equation}
E_{\text{FC,e}} = E_\text{g}^\text{*}- E_\text{e}^\text{*}.
\end{equation}
Then the photon emission proceeds, with energy
\begin{equation}
E_{\text{em}} = E_\text{e}^\text{*} - E_\text{e}, 
\end{equation} 
and the system relaxes into the electronic ground state, with a release of energy given by the Franck-Condon shift in the ground state
\begin{equation}
E_{\text{FC,g}} = E_\text{e} -  E_\text{g}.
\end{equation}
The two Franck-Condon shifts combine to give the observable Stokes shift
\begin{equation}
\Delta S = (E_\text{g}^\text{*} - E_\text{g}) - (E_\text{e}^\text{*} - E_\text{e}).
\end{equation}
Sometimes, experiments yield also the zero-phonon line that corresponds to
\begin{equation}
E_{\text{ZPL}} = E_\text{e}^\text{*}  - E_\text{g}.
\end{equation}

The calculation of the zero-point motion from first principles is available inside the ABINIT software~\cite{Ponce2014a,Ponce2015} but the effect is small, 
and computationally expensive for such materials and therefore has been left for further study.   
 
These semi-classical absorption and emission energies, as well as the Stokes shift, can be directly compared with experimental data and can be used
to identify the luminescence site.

This approach can also provide other quantities of interest without performing additional first-principle calculations. 
Indeed, by assuming parabolicity it is possible to extract the 1D-coordinate of a crossing point (if any), and thus, 
an estimate (upper bound) for the energy barrier needed to have $4f-5d$ crossover.\cite{toyozawa2003optical} This in turn gives some information on the likelihood of a $4f-5d$ non-radiative recombination.
In fact, there is a debate about the thermal quenching mechanism for the luminescence of RE ion doped phosphors. Two mechanisms are often invoked: the auto-ionization and the $4f-5d$ crossover.\cite{samuel2016}
The corresponding energy barriers for the two processes are E$_{\text{dC}}$ and E$_{\text{fd}}$, respectively, as shown in Fig.~\ref{figure_2}. 
Following the 1D-CCD model, the geometry for the $4f-5d$ crossover, Q$_{\text{fd}}$, would be the linear combination of the Q$_{\text{g}}$ and Q$_{\text{e}}$ geometries. As a result, we define Q$_{\text{fd}}$ as:

\begin{equation}
Q_{\text{fd}} = (1-x)Q_{\text{g}} + xQ_{\text{e}}. 
\end{equation} 

The curvatures of the ground-state and excited-state parabola are directly deduced from the Franck-Condon shifts. They might be different, and we define a parameter to show 
the difference of curvatures of ground and excited state, $\Delta$C as follows:
\begin{equation}
\Delta C = (E_\text{g}^\text{*} - E_\text{e}^\text{*}) - (E_\text{e} - E_\text{g}) = E_{\text{FC,e}} - E_{\text{FC,g}}.
\end{equation} 

Then, solving the second-degree equation that defines the crossing point, one gets \textit{1/x} as:
 \begin{equation}
1/x = [E_{\text{FC,e}} + \sqrt{E_{\text{FC,e}}^2 - E_{\text{abs}}\Delta C}]/E_{\text{abs}}.
\end{equation} 

The thermal quenching barrier, E$_{\text{fd}}$ can be finally determined by:
\begin{equation}
E_{\text{fd}} = E_{\text{FC,e}}(x-1)^2. 
\end{equation} 

This result reduces to 
\begin{equation}
E_{\text{fd}} = \dfrac{{E_{\text{em}}}^2}{4E_{\text{FC,e}}} 
\end{equation}
when the curvatures of ground and excited state are identical ($\Delta$C = 0), which is also the result of a recent work.\cite{4f-5d}

Beside the information on the thermal energy barrier for the $4f-5d$ crossover, the 1D-CCD model can also be used to calculate approximately the
main characteristics of the luminescence spectrum line, as shown 
for some doped semiconductors, MgO, ZnO and GaN, with a nice agreement between experiment and theory.\cite{1d-ccd}  
At variance with doped semiconductors, the optical transitions of RE-doped phosphors have an intra-atomic $4f\leftrightarrow5d$ characteristic,
with a ground state or excited state that is more localized than the transition in such semiconductors.
Thus, it is worth to test the outcome of such 1D-CCD model for the luminescence spectrum line shape of RE-doped phosphors. 

For this purpose, one supposes that the large number of vibrational modes contributing to the line shape can be simplified into one effective vibrational mode. 
The parameters entering the 1D model are the normal coordinate Q (connecting Q$_{\text{g}}$ and Q$_{\text{e}}$), the displacement of the nuclei at the potential energy minimum $\Delta$R, 
the modal mass M of the effective vibration, and the effective vibrational frequencies $\Omega_{\text{g}}$ and $\Omega_{\text{e}}$. These parameters can be calculated as follows: \cite{1d-ccd}
\begin{align}
\Delta Q^2 &=\sum_{\alpha, \text{i}}m_{\alpha}(R_{\alpha \text{i},\text{e}}-R_{\alpha \text{i},\text{g}})^2\\
\Delta R^2 &=\sum_{\alpha, \text{i}}(R_{\alpha \text{i},\text{e}}-R_{\alpha \text{i},\text{g}})^2 \\
M &= \Delta Q^2/\Delta R^2, 
\end{align}  
where $\alpha$ denotes atoms in the supercell calculation, $i$ denotes the Cartesian directions, m$_{\alpha}$ is the mass of atom $\alpha$ and the R$_{\alpha i, g(e)}$ are 
the atomic coordinates in the ground and excited states, respectively. 
The modal mass M is an average of the masses of the ions involved in the displacement, weighted by the square of the nuclei displacements.

Based on the effective vibrational model and the total energy obtained through the ${\Delta}$SCF method, the effective vibration frequencies are:
\begin{equation}\label{Omg}
\Omega_\text{g}^2 = 2E_{\text{FC,g}}/\Delta Q^2,
\end{equation}
\begin{equation}\label{Ome}
\Omega_\text{e}^2 = 2E_{\text{FC,e}}/\Delta Q^2.
\end{equation} 

Then, the Huang-Rhys factors that denote the average numbers of phonons emitted in the ground- and excited-state geometry can be obtained as: 
\begin{equation}\label{Sabs}
S_{\text{abs}} = E_{\text{FC,e}}/(\displaystyle \hbar \Omega_\text{e}),
\end{equation}
\begin{equation}\label{Sem}
S_{\text{em}} = E_{\text{FC,g}}/(\displaystyle \hbar \Omega_\text{g}).
\end{equation}

Treated quantum mechanically, the harmonic oscillator in the excited state will yield a series of quantized energy levels, and their corresponding wavefunctions.
The luminescence emission spectrum line shape at 0~K can be expressed using matrix elements of the transition, as:\cite{book1,GaN}
\begin{equation}\label{lumin}
L(\displaystyle \hbar \Omega) = I_0  \sum_{n}\dfrac{e^{-S_{\text{em}}}S_{\text{em}}^{n}}{n!}\delta(E_{\text{ZPL}}-n\hbar\Omega_\text{g} -\hbar\Omega),
\end{equation}  
where I$_0$ is a normalization factor, $E_{\text{ZPL}}$ is the energy of the zero-phonon line and $\delta$ is a broadened Dirac function. 

To treat Eq.~\eqref{lumin}, a semi-classical approximation can be used to describe the expectation value of $Q$ and compute the
density of transition as a function of the emitted energy.
From there, the FWHM of the emission peak, $W$ at 0~K can be calculated as:\cite{book1,GaN} 
\begin{equation}
\label{W0}
W(0) = S_{\text{em}} \hbar \Omega_\text{g}\sqrt{8\ln2}/\sqrt{S_{\text{abs}}}.
\end{equation} 

At temperature $T$, the FWHM can be expressed as
\begin{equation}\label{lumine}
W(T) = W(0)\sqrt{\coth(\hbar \Omega_\text{e}/2 k_BT)},
\end{equation} 
where $k_B$ is the Boltzmann constant.

Also, the energy shift of the emission peak with temperature can be calculated as:\cite{temperture,GaN}
\begin{multline}\label{tem}
E_{\text{em}}(T) - E_{\text{em}}(0) =\\
 \Big(\dfrac{\Omega_\text{g}^2 - \Omega_\text{e}^2}{\Omega_\text{e}^2}+\dfrac{8\Omega_\text{g}^4 \Delta S(0)}{\Omega_\text{e}^2(\Omega_\text{g}^2+\Omega_\text{e}^2)E_{\text{em}}(0)}\Big) k_\text{B}T.
\end{multline}

In Eqs. (\ref{Omg})-(\ref{tem}), we consider the general case in which the effective phonon frequency of ground and excited states can be different. 
If $\Omega_{\text{g}}$ is equal to $\Omega_{\text{e}}$, Eqs.~\eqref{W0}-\eqref{lumine} are the same as the expressions presented in Ref.~[\onlinecite{4f-5d}].

Somehow, Our analysis will actually need to improve on  
Eqs. (\ref{W0}-\ref{tem}). Indeed these equations assume that the harmonic approximation is valid for the entire range of $Q$ values, from $Q_g$ to $Q_e$, for both the ground state and the excited state, and afterwards linearize the behaviour of the ground state around $Q_e$.  Instead, we will find later
that while the harmonic approximation is valid for the ground state, in some cases the excited state energy is
not well represented by a parabola. Actually, one can reformulate Eqs. (\ref{W0}) to use
the harmonic approximation only in a neighborhood of $Q_e$ for the excited state, thus
rely on the curvature at $Q_e$,
and to use of the local slope of the ground-state energy curve at $Q_e$, instead of relying
on an data that involves information over the full $Q_g$ to $Q_e$ range, assuming harmonicity.
In this context, Eqs.(\ref{Omg})-\ref{Sem}) are not valid anymore. 

We reformulate Eq.(\ref{W0}) as follows. The full width at half maximum of the probability density 
of the lowest state of a one-dimensional harmonic oscillator with potential energy $E(R)$ and mass $M$, $W_{PD}$, is given by 
\begin{equation}\label{FWHM_PD}
W_{PD}=2 \sqrt{\ln 2} \left( \frac{\hbar}{M \Omega} \right)^{1/2}.
\end{equation}
The oscillator frequency $\Omega$ is directly linked to the curvature of the potential energy expressed
as a function of the normal coordinate $Q=M^{1/2}R$:
\begin{equation}
\Omega= \left( \frac{\partial^2 E}{\partial Q^2} \right)^{1/2}.
\label{OmeCurvature}
\end{equation}
These equations will be used for the spread of the probability to find the excited state around  $Q_e$. 
The semi-classical approximation is completed by supposing that the spread of the probability in $R$
translates linearly to a spread in energy for the emission, by way of the linear slope of $E_g(R)$ evaluated
at $Q_e$. Thus, the following equation replaces Eq.(\ref{W0}),
\begin{equation}
W(0)=2 \sqrt{\ln 2} \left( \frac{\hbar}{\Omega_e} \right)^{1/2} \left( \frac{\partial E_g}{\partial Q} \right),
\end{equation}
where only local information around $Q_e$ are used.

If the harmonic approximation is valid for the ground-state energy $E_g$ 
in the entire range $Q_g$ to $Q_e$, its slope is
\begin{equation}
\frac{\partial E_g}{\partial Q}=\frac{2E_{FC,g}}{\Delta Q},
\end{equation}
and Eqs.(\ref{Omg}) and (\ref{Sem}) are valid again, giving

\begin{equation}
W(0) = S_{\text{em}} \hbar \Omega_\text{g}\sqrt{8\ln2} \left( \frac{\hbar}{\Omega_e} \right)^{1/2}
   \frac{1}{\Delta Q}
\label{Eq27}
\end{equation} 
Of course, this equation reduces to Eq.(\ref{W0}) in case the excited state energy $E_e$ is harmonic,
or equivalently, if Eqs.(\ref{Ome}) and (\ref{Sabs}) are valid.

In section II.C, in order to analyze the main effect of the anharmonicity of the excited state on $W(0)$,  
we will rely on the proportionality of $W(0)$ to $\left( \frac{\partial^2 E}{\partial Q^2} \right)^{-1/4}$
evaluated at $Q_e$, which is a consequence of Eqs.(\ref{Eq27}) and (\ref{OmeCurvature}) at constant $\Delta Q$ and of the validity of the harmonic approximation for the ground state.

\subsection{Dorenbos' Semi-empirical model}
\label{Dorenbos}
\begin{figure}[h]
\includegraphics[scale=0.35]{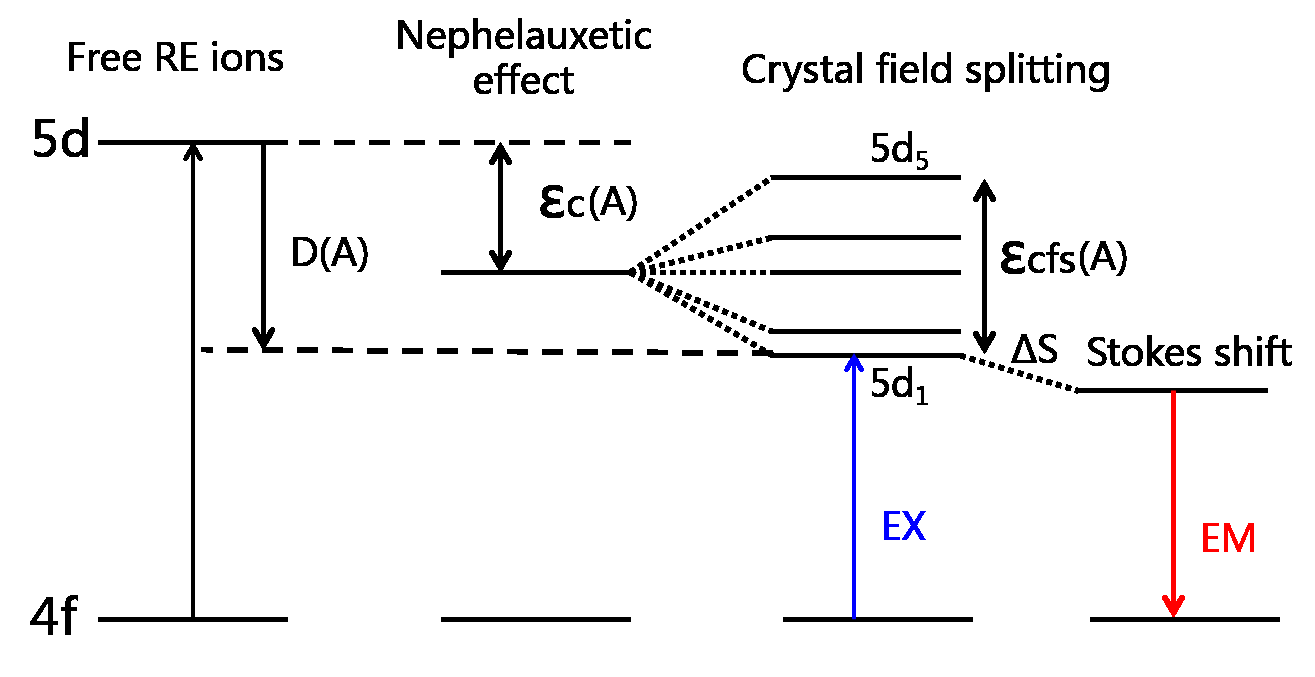}
\caption{Dorenbos' semi-empirical model of the $4f-5d$ transition energy of the Eu$^{2+}$ ion. The $\varepsilon_{c}$(\textit{A}), $\varepsilon_{cfs}$(\textit{A}) and \textit{D}(\textit{A}) indicate the centroid shift, crystal field splitting and redshift of Eu$_{5d}$ energy level in compound \textit{A}, respectively. The energy level of RE ions is aligned to the $4f$ states.}\label{figure_3}
\end{figure}

To assess our first-principle calculations of the transition energies and Stokes shift, we also studied the Dorenbos' semi-empirical model 
for the absorption and emission processes in Eu$^{2+}$-doped materials (see Fig.~\ref{figure_3}).
At present, quantitative expressions for the energy of the first allowed $4f\rightarrow5d$ transition of the Ce$^{3+}$ ion in inorganic materials have been proposed by Dorenbos, for selected anions.\cite{dorenbos1,dorenbos2,dorenbos4}

Following Dorenbos' semi-empirical model, the energy of the first allowed $4f\rightarrow5d$ transition of the free RE ion is lowered by the crystalline environment, with a shift denoted \textit{D}(\textit{A}). 
This lowering is the sum of the spectroscopic redshift arising from the centroid shift of the RE$_{\textit{5d}}$ energy, $\varepsilon_{c}$(\textit{A}), 
and the crystal-field splitting, $\varepsilon_{cfs}$(\textit{A}), of the RE$_{\textit{5d}}$ states. For the Ce$^{3+}$ ion, the redshift $D(A)$ can be written as \cite{dorenbos1}
\begin{equation}
D(A) = \varepsilon_{\textit{c}}(A) + \frac{\varepsilon_{\textit{cfs}}(A)}{r(A)} - 1890\,\text{cm}^{-1},
\end{equation}
in which the $\varepsilon_{c}$(\textit{A}) is the centroid shift of the Ce$_{5d}$ energy relative to the free ion, defined as follows: 
\begin{equation}
\varepsilon_{c}(A) = 1.44\times{10}^{17}\sum_{\text{i=1}}^N\dfrac{\alpha_{\textit{sp}}^i}{R_\textit{i}^6}.
\end{equation}  

In the above formula, $\alpha_{sp}^i$ is the spectroscopic polarization of anion $i$ located at distance R$_i$ from the Ce$^{3+}$ ion in the relaxed structure. 
The summation is over all anions $N$ in the coordination environment of RE ions.  At present, quantitative relationships between $\alpha_{sp}^i$ 
and the electronegativity of the cations for oxides, nitrides and fluorides, have been proposed as:\cite{Wang2015}
\begin{align}
\alpha_{sp}^O &= 0.33 + \dfrac{4.8}{\chi_{\text{av}}^2},\\
\alpha_{sp}^N &= 0.87 + \dfrac{18.76}{\chi_{\text{av}}^2},\\
\alpha_{sp}^F &= 0.15 + \dfrac{0.96}{\chi_{\text{av}}^2},
\end{align}
where the electronegativity is 
\begin{equation}
\chi_{av} = \dfrac{1}{N}\sum_{\text{i=1}}^M \frac{Z_{\text{i}}\chi_{\text{i}}}{\gamma}.
\end{equation}   

This formula is rationalized by considering that a cation of charge Z$_{i}$ will bind on average with Z$_{i}$/$\gamma$ anions of charge -$\gamma$. 
The summation is over all cations $M$ in the compound, and $N$ is the number of anions.\citep{Wang2015} 

Another parameter affecting the spectroscopic redshift is the contribution from the crystal field shift, $\frac{1}{r(A)}\varepsilon_{cfs}(A)$. 
The crystal-field splitting $\varepsilon_{cfs}(A)$ is defined as the energy difference between the lowest and highest $5d$ levels. 
A fraction 1/r(\textit{A}) contributes to the redshift, where r(\textit{A}) usually varies between 1.7 and 2.4. The $\varepsilon_{\text{cfs}}(A)$ is determined as: 
\begin{equation}
\varepsilon_{cfs} = \frac{\beta}{R_{\textit{av}}^2}.
\end{equation}     

Here, $\beta$ is a parameter related to the shape and size of the anion polyhedron coordinated to the Ce$^{3+}$ ion, 
and $R_{\text{av}}$ is the average distance between the Ce$^{3+}$ ion and anions in the relaxed structure. 
Based on $D(A)$ and the energy of the first $4f\rightarrow5d$ transition of Ce$^{3+}$ as a free (gaseous) ion, 49340 cm$^{-1}$, 
the transition energy of Ce$^{3+}$ ion in compound A can be calculated as:\cite{jia2016,jia2017}
\begin{equation}
E(A) = 49340\,\text{cm}^{-1} - D(A). 
\end{equation}

Compared to the comprehensive study of Ce$^{3+}$ ion in inorganic compounds, the related work for the Eu$^{2+}$ doped phosphors is limited 
because of the more complex electronic configuration of the Eu$^{2+}$ ion.
However, a similar quantitative expression for the Eu$^{2+}$ ion can be expected since these two ions have similar $4f\rightarrow5d$ neutral excitations. 
A semi-empirical relationship between the redshift for the $5d$ state of Eu$^{2+}$ and Ce$^{3+}$ ions, in the same host, has been determined:\cite{dorenbos5}

\begin{equation}\label{Deq}
D(Eu^{2+}, A) = 0.64\times D(Ce^{3+}, A) - 0.233 \,\text{eV}. 
\end{equation}

This equation indicates that the redshift of the $5d$ state of the Eu$^{2+}$ ion is smaller than the one of the Ce$^{3+}$ ion. 
The difference in slope by a factor 0.64 might be due to the effect of the remaining six Eu$_{4f}$ electrons in the excited state 
and the intercept of 0.233 might be due to the different ionic radius of the two ions. 
Using Eq.~\eqref{Deq}, we propose a direct expression for the Eu$^{2+}$ ion in compound \textit{A}: 
\begin{equation}
D(Eu^{2+}, A) = \varepsilon_{\textit{c}}(Eu^{2+}, A) + \dfrac{\varepsilon_{\textit{cfs}}(Eu^{2+}, A)}{r(Eu^{2+}, A)} - 1890\,\text{cm}^{-1},
\end{equation} 
where 
\begin{align}
\varepsilon_{\text{c}}(Eu^{2+}, A) &= 1.44 \times 0.64\times{10}^{17}\sum_{\text{i=1}}^N\dfrac{\alpha_{\textit{sp}}^i}{R_i^6},\\
\varepsilon_{\text{cfs}} &= 0.64 \times \dfrac{\beta}{R_{\textit{av}}^2}.
\end{align}  

Additionally, we assume that the fitting formulation for $\alpha_{\text{sp}}$ and the value of $\beta$ from the study of Ce$^{3+}$-doped materials 
are still valid for the Eu$^{2+}$-doped ones. Then, the redshift of Eu$_{\textit{5d}}$ state can be determined with available first-principles geometry information on the ground and excited states. 
Finally, the transition energy of Eu$^{2+}$ ion in compound \textit{A} is:
\begin{equation}
E(Eu^{2+}, A) = 34004\,\text{cm}^{-1} - D(Eu^{2+}, A), 
\end{equation}
where the first $4f\rightarrow5d$ transition energy of 34004 cm$^{-1}$ for the free Eu$^{2+}$ ion is from Ref.~[\onlinecite{dorenbos5}].

\section{RESULTS AND DISCUSSION}
\label{validation and usage} 

We present successively, for the set of fifteen representative Eu$^{2+}$-doped materials: (1) absorption and emission energies, and associated Stokes shifts; (2) energy barrier for $4f-5d$ crossover; (3) FWHM of the emission peak, and also the temperature shift; and (4) the extension of Dorenbos' semi-empirical model to Eu$^{2+}$-doped materials.

\subsection{Absorption and emission energies, and associated Stokes shift.}
\label{validation}

Table \ref{table-1} lists the absorption, emission energy and Stokes shift for the fifteen representative Eu$^{2+}$-doped materials from our first-principles calculations and  experiment. This table has eighteen entries, as for the fifteen materials, two cases are distinguished: for SrAl$_2$O$_4$ and SrLiAl$_3$N$_4$, two different substitutional sites for Eu are considered, while two different atomic geometries are considered for CaAlSiN$_3$, as discussed in the Appendix.

\begin{figure*}
\includegraphics[scale=0.25]{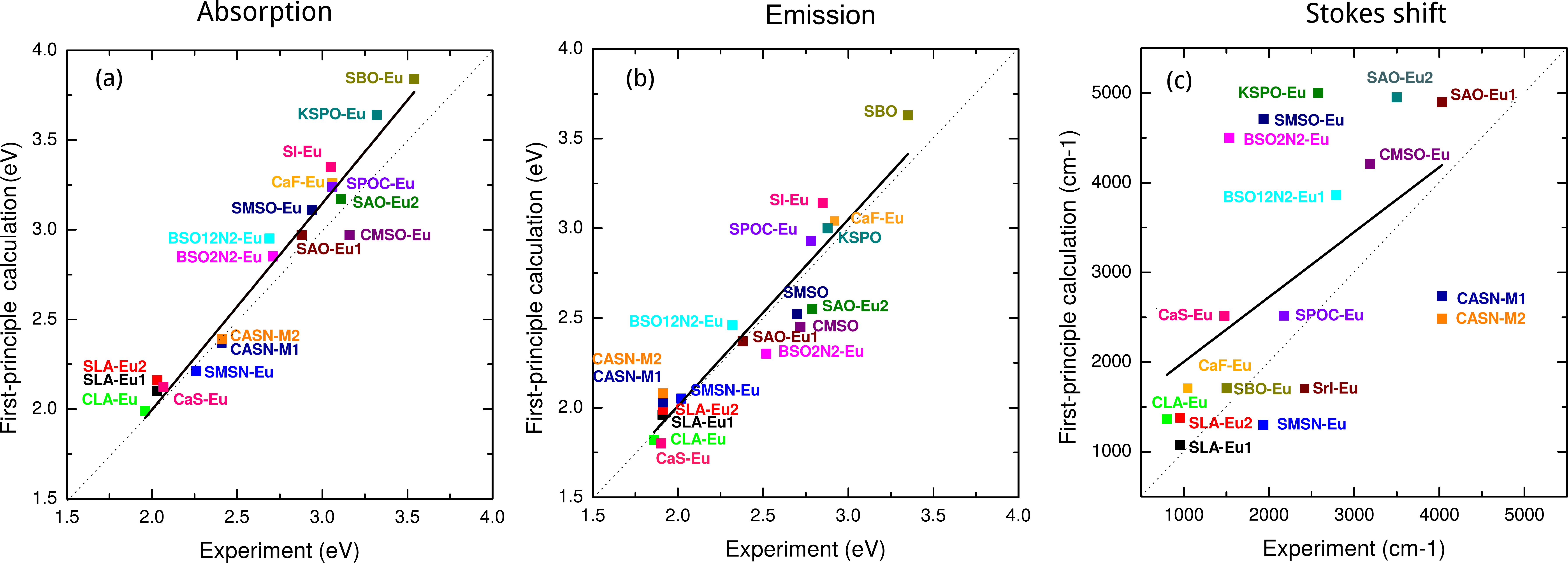}
\caption{Comparison between experimental results and first-principles calculations: (a) absorption energy, (b) emission energy, and (c) Stokes shift. The black line is the least-square fit of the data.}\label{figure_4}
\end{figure*}

For SrLiAl$_3$N$_4$:Eu, a ground-state study has already been performed in Ref.~\onlinecite{SLA-2}. To show the difference in the electronic band structure of ground and excited state using the $\Delta$SCF method, a more detailed presentation has been given in the Appendix, including electronic structure plots.
Similar data have been computed for all materials and can be obtained upon request to the corresponding author.  
Fig.~\ref{figure_4} shows the comparison between theory and experiment leading to a mean relative error (MRE) of 3.8\% and 2.3\% for the absorption and emission energies, respectively (see Table~\ref{table-2} for more information). 
The Stokes shift is a much more sensitive quantity, resulting in a 40\% mean relative error (MRE) with respect to experiment.

The calculated transition energies match experiment within 0.3~eV for all the fifteen materials. 
Moreover, the slope of the fitting line for the absorption and emission energy is quite close to unity which
indicates a good predictive capability of the $\Delta$SCF method for the transition energies of the Eu$^{2+}$-doped phosphors. 

The obtained Stokes shifts give an error of about 20\% in general. However, much larger errors are obtained
for the eight cases shown in bold in Table~\ref{table-1}, which are less satisfactory. The origin of these larger errors might be due to the cation disorder in the crystal structure or to the inaccurate assessment of the absorption 
and emission spectra, but might also indicate an intrinsic limitation of the theoretical approach, which provides usually an accuracy of 0.3 eV for the transition energy. The Stokes shift arises from a modification of the local geometry around the Eu ion upon electronic excitation. Indeed the strongly localized \textit{4f} state is replaced by a more delocalized state with \textit{5d} dominance, inducing less screening of the ion positive charge, and thus greater attraction of the neighboring anions (e.g. Oxygen or Nitrogen). 
The local environment of the Eu$^{2+}$ ions in these eight compounds was checked to detect the possible  origin of the large error in the Stokes shift. Unfortunately, we saw no obvious relationship between the eight compounds from the analysis of their coordination number, crystal environment and bond length.

In addition to the fifteen Eu$^{2+}$-doped materials, we also studied the case of Eu$^{2+}$ ion in the CaO host. 
In the excited state band structure with the ground-state geometry, the occupied Eu$_{\textit{5d}}$ state was not located inside the band gap, 
which indicates a non-luminescent character of the Eu$^{2+}$ ion in this host. Also, with the same value of U used for the other materials, the seven $4f$ states entered the valence band. Reducing the value of U succeeded in correcting this failure, but did not lead to a $5d$ state inside the band gap. This result is in opposition to the experimental observations. 
At present, the reason for this failure of the $\Delta$SCF method is not clear. It might be due to the difficulty in finding the global energy minimum of the system, and/or the small DFT+U energy gap of the CaO:Eu. 
A more careful study should be conducted based on a higher-level computational methodology such as
a hybrid functional or the GW method. Indeed one would obtain a better starting electronic structure, 
and consequently describe luminescence of Eu$^{2+}$ ion in this host.
Taking into account such information, it can be concluded that the present study does not guarantee that the $\Delta$SCF method would work well for every Eu-doped compound. 
Still, the general agreement for most of compounds encourages the use of the $\Delta$SCF method based on the PBE exchange-correlation functional, 
provided that the Eu$_{4f}$ and Eu$_{5d}$ states are found in the band gap. 

\begin{table}[h]
\centering \renewcommand\arraystretch{1.5}
\caption{Absorption/emission energy (eV) and Stokes shift (cm$^{-1}$), from first-principles calculations as well as from experiment. The numbers in bold deviate by more than 1000 cm$^{-1}$ from experiment. The notation for the Eu sites in the SrAl$_2$O$_4$:Eu and Sr[LiAl$_3$N$_4$]:Eu cases are from the references [\onlinecite{SAO}] and [\onlinecite{SLA}] and shown in Figure~\ref{figure_a1} and Figure~\ref{figure_a3}, respectively. The results of Ba$_3$Si$_6$O$_{12}$N$_2$:Eu and Sr$_5$(PO$_4$)$_3$Cl:Eu are obtained from the Eu$^{2+}$ ions in the confirmed luminescence sites.\cite{samuel2016,SPOC}. More detailed information can be found in the text. } \label{table-1} 
\begin{tabular}{l|ccc|cccc} 
\hline\hline
 Compound &\multicolumn{3}{c|}{Calculation}&\multicolumn{4}{c}{Experiment} \\
 & Abs &Em &$\Delta$S &Abs &Em &$\Delta$S & Ref. \\ \hline
 SrB$_4$O$_7$:Eu & 3.845 & 3.633 & 1710 & 3.54 & 3.35 & 1502 & [\onlinecite{SBO}] \\
 KSrPO$_4$:Eu & 3.612 & 2.998 & \textbf{4920} & 3.32 & 2.88 & 3500 & [\onlinecite{KSPO}]\\ 
 CaMgSi$_2$O$_6$:Eu & 2.969 & 2.447 & \textbf{4218} & 3.16 & 2.72 & 3188 & [\onlinecite{CMSO}]\\  
 SrAl$_2$O$_4$:Eu2 & 3.17 & 2.547 & \textbf{4996} & 3.11 & 2.79 & 2581 & [\onlinecite{SAO}]\\  
 Sr$_5$(PO$_4$)$_3$Cl:Eu & 3.238 & 2.926 & 2516 & 3.06 & 2.78 & 2178 & [\onlinecite{SPOC}]\\ 
 CaF$_2$:Eu & 3.257 & 3.045 & 1774 & 3.06 & 2.92 & 1047 & [\onlinecite{CaF2}]\\
 SrI$_2$:Eu & 3.349 & 3.138 & 2339 & 3.05 & 2.85 & 2420 & [\onlinecite{SrI2}]\\ 
 Sr$_2$MgSi$_2$O$_7$:Eu & 3.107 & 2.523 & \textbf{4726} & 2.94 & 2.70 & 1936 & [\onlinecite{SMSO}]\\ 
 SrAl$_2$O$_4$:Eu1 & 2.968 & 2.361 & 4839 & 2.88 & 2.38 & 4033 & [\onlinecite{SAO}]\\ 
 BaSi$_2$O$_2$N$_2$:Eu & 2.855 & 2.297 & \textbf{4436} & 2.71 & 2.52 & 1532 & [\onlinecite{BSO2N2}]\\ 
 Ba$_3$Si$_6$O$_{12}$N$_2$:Eu & 2.940 & 2.461 & 3952 & 2.69 & 2.32 & 2790 & [\onlinecite{BSO12N2}]\\
 CaAlSiN$_3$:Eu,M-I & 2.367 & 2.028 & \textbf{2742} & 2.41 & 1.91 & 4032 & [\onlinecite{CASN}]\\
 CaAlSiN$_3$:Eu,M-II & 2.387 & 2.079 & \textbf{2508} & 2.41 & 1.91 & 4032 & [\onlinecite{CASN}]\\ 
 Sr[Mg$_3$SiN$_4$]:Eu  & 2.216 & 2.055 & 1290 & 2.26 & 2.02 & 1935 & [\onlinecite{SMSN}]\\ 
 CaS:Eu & 2.120 & 1.810 & \textbf{2500} & 2.07 & 1.90 & 1466 & [\onlinecite{CaS}]\\
 Sr[LiAl$_3$N$_4$]:Eu1 & 2.095 & 1.962 & 1129 & 2.03 & 1.91 & 956 & [\onlinecite{SLA}]\\
 Sr[LiAl$_3$N$_4$]:Eu2 & 2.160 & 1.989 & 1371 & 2.03 & 1.91 & 956 & [\onlinecite{SLA}]\\
 Ca[LiAl$_3$N$_4$]:Eu  & 1.992 & 1.823 & 1371 & 1.96 & 1.86 & 800 & [\onlinecite{CLA}]\\ 
\hline\hline 
\end{tabular} 
\end{table} 

\begin{table}
\centering
\renewcommand\arraystretch{1.5}
\caption{Statistical analysis of transition energies (eV), Stokes shifts $\Delta$S (cm$^{-1}$) and FWHM (eV) from first-principles calculations. 
ME, MAE, MRE and MARE stand for the mean error, mean absolute error, mean relative error, and mean absolute relative error, respectively. 
The slope, intercept and coefficient of determination (R-Square) correspond to the least-square fitting lines 
shown in Fig.~\ref{figure_4} for the transition energies and Stokes shift, and in Fig.~\ref{figure_5} for the FWHM. }
\label{table-2}
\begin{tabular}{l|cccc}
\hline\hline
 &\multicolumn{4}{c}{First-principles calculation} \\
 & Absorption & Emission& Stokes shift & FWHM  \\ \hline
ME   & 0.111 & 0.058 & 651 & 0.023  \\
MAE  & 0.144 & 0.159 & 1118 & 0.052  \\
MRE (\%) & 3.83 & 2.27 & 40.06 & 16.7 \\
MARE (\%)& 5.03 & 6.36 & 54.85 & 30.03 \\
Slope   & 1.127 &1.040 & 0.723 & 0.584 \\
Intercept&-0.233 & -0.069 & 1279 & 0.111 \\ 
R-Square (\%)& 95.0 & 88.6 & 26.1 & 21.9 \\ \hline\hline
\end{tabular}
\end{table}

Going beyond the tabulation of results, we emphasize that the $\Delta$SCF method can be a useful tool for the identification of luminescent center(s) 
in Eu$^{2+}$-doped phosphors. In the present study, four materials (Sr[LiAl$_3$N$_4$], Ba$_3$Si$_6$O$_{12}$N$_2$, SrAl$_2$O$_4$ and Sr$_5$(PO$_4$)$_3$Cl) 
among the list of fifteen materials have at least two possible substitution sites. 
For the Sr[LiAl$_3$N$_4$]:Eu, the $\Delta$SCF method indeed shows that the Eu$^{2+}$ ions at the two inequivalent Sr$^{2+}$ sites give a rather similar optical behavior, 
which correlates well with the very narrow emission band in this red phosphor.
This is explained by the very similar local environment of the two sites, both in terms of coordination (the Eu ion is placed inside a slightly distorted cube in both cases) 
and neighbour distances. More details are given in the Appendix, see e.g. Fig.~\ref{figure_a3}.

In contrast, experiment shows two emission peaks in SrAl$_2$O$_4$:Eu$^{2+}$. 
The high-energy emission peak is centered at 435 nm while the lower one is at around 500 nm. We indeed obtain theoretically two well-separated
absorption and emission energies.
The lower emission peak 
is from Eu$^{2+}$ ion at the Sr$^{2+}$ site which gives a perfect linear chain along the \textit{z} direction of the SrAl$_2$O$_4$ crystal. 
This result is consistent with the empirical assessment from experiment.\cite{SAO}

For Ba$_3$Si$_6$O$_{12}$N$_2$:Eu$^{2+}$ and Sr$_5$(PO$_4$)$_3$Cl:Eu$^{2+}$, the existing experimental works indicate that there is only one luminescent center. 
Our previous theoretical result on Ba$_3$Si$_6$O$_{12}$N$_2$:Eu$^{2+}$ has determined that in this phosphor the luminescent center arises from the Eu$^{2+}$ ion located on the Ba$^{2+}$ 
site that is coordinated with six oxygens.\cite{samuel2016} 
This conclusion has been further confirmed in this work. 
The results listed in Table~\ref{table-1} corresponds to this site and matches the experimental data. 
For Sr$_5$(PO$_4$)$_3$Cl, the coordination environments of the two Sr$^{2+}$ sites are quite different: 
one is coordinated with Cl atoms (C$_{1h}$ point group) and the other is not coordinated with Cl atoms (C$_3$ point group).
Experimentalists have postulated that the luminescent center originates from the Eu$^{2+}$ ion at the Sr (C$_3$ point group) site.\cite{SPOC} This idea is validated by our $\Delta$SCF method.

Beside the contribution of the $\Delta$SCF method to the identification of the luminescent centers, results for CaAlSiN$_3$:Eu$^{2+}$, which is a commercial red emission phosphor 
with a broad band, are also quite interesting. Indeed, some disorder between Al$^{3+}$ and Si$^{4+}$ cations is present in the crystal structure. 
Mikami and coworkers\cite{CASN1} have proposed two ordered structures to mimic the real crystal. The corresponding structural properties indeed show the plausibility of the two ordered structures. 
Here, we validate this idea based on the luminescence of Eu$^{2+}$ ion in the two ordered crystal structures of CaAlSiN$_3$. 
The consistency of the calculated transition energies and Stokes shifts with experiment gives a firm interpretation. The two ordered crystal structures of CaAlSiN$_3$ are shown in the Appendix.    


\subsection{Energy barrier for $4f-5d$ crossover}
\label{4f-5d}

Thermal quenching behaviour is an important parameter in RE-doped phosphors, especially for high-power LED applications. 
The ratio between non-radiative recombinations and radiative ones increases with temperature.
There is a debate in the literature about the dominant non-radiative mechanism.
Two mechanisms have been proposed: the auto-ionization process and the $4f-5d$ crossover. 
In the former mechanism, the localized RE$_{5d}$ electron has an increasing probability to be promoted to the conduction band minimum of the host with increasing temperature, 
so becoming delocalized and followed by non-radiative recombination happening at some defect sites in the host. 
The energy jump for the auto-ionization process of the RE$_{5d}$ electron is E$_{\text{dC}}$, as shown in Fig.~\ref{figure_2}. 
At variance, the thermal quenching via the $4f-5d$ crossover considers that the localized RE$_{5d}$ electron can be transferred to a RE$_{4f}$ state, in a highly excited vibrational state,
after which the energy is released through a multiphonon process. A good indicator of the likelihood of this process is given
by the energy barrier needed to reach the crosspoint of the energy potential surface of RE$_{5d}$ and  RE$_{4f}$ states, in the 1D-CCD model. 
The corresponding energy barrier is indicated as E$_{\text{fd}}$ in Fig.~\ref{figure_2}. 

In the literature, following a simple semi-classical argument, a criterion has been proposed to quantify the possibility that, immediately after the excitation, 
the electron would have sufficient energy to propagate straightforwardly to the crossing point. 
Indeed, if $E_{\text{fd}}$ is smaller than $E_{\text{FC,e}}=E_\text{g}^*-E_\text{e}^*$, the excited state Eu$_{5d}$ electron would immediately have the opportunity to reach the $4f-5d$ crossing point.
This is often formulated by the following criterion for immediate non-radiative recombination, in terms of a parameter $\Lambda$ \cite{book1,book2}:
\begin{equation}
\Lambda = \dfrac{E_{\text{FC,e}}}{E_{\text{abs}}} > 0.25.
\end{equation}
This expression reduces to the comparison of E$_{\text{fd}}$ and E$_{\text{FC,e}}$, provided that the curvatures are equal.
For lower values of $\Lambda$, immediate thermal quenching by the $4f-5d$ crossover mechanism is less likely, and both the radiative recombination and the auto-ionization mechanism will be competing.

In this section, we evaluate E$_{\text{fd}}$ for the Eu$^{2+}$-doped phosphors.
The theoretical method for the calculation of the E$_{\text{fd}}$ was explained in Section~\ref{Dorenbos}.
Note the role of the $\Delta$C parameter that might be such that the 4f and 5d states do not even cross.

Table~\ref{table-3} lists the first-principles parameters as well as the $\Lambda$ parameter and E$_{fd}$ results.  
The largest $\Lambda$ parameter is less than 0.1, indicating the low probability of immediate $4f-5d$ non-radiative recombination.
All the values of E$_{\text{fd}}$ are above 1.5~eV, which is much larger than the experimental energy barrier 
of thermal quenching, usually found around 0.5~eV. Of course, such estimate is done within the 1D-CCD, which weakens our analysis.
Still, these results indicate that the mechanism of $4f-5d$ crossover is likely not the major mechanism for the thermal quenching behaviour of the Eu$^{2+}$-doped phosphor. 
We therefore infer that the auto-ionization process should be the dominant thermal quenching mechanism.

\begin{table*}
\centering
\renewcommand\arraystretch{1.5}
\caption{Estimation of the energy barrier E$_{\text{fd}}$ (eV), and related data, for the $4f-5d$ crossover, in the fifteen Eu$^{2+}$-doped materials. See text for the corresponding definitions. The value `--' for x indicates that the 4f and 5d curves do not cross in the parabolic approximation.}
\label{table-3}
\begin{tabular}{lcccccccc}\hline
Compound                      & E$_{\text{abs}}$ & E$_{\text{ZPL}}$ & E$_{\text{FC},\text{g}}$ & E$_{\text{FC},\text{e}}$ & $\Lambda$ & $\Delta$C & x       & E$_{\text{fd}}$    \\\hline  
SrB$_4$O$_7$:Eu                   & 3.845 & 3.736 & 0.103 & 0.109 & 0.028 & 0.005     & -- &    $\infty$     \\
KSrPO$_4$:Eu                  & 3.621 & 3.300 & 0.302 & 0.321 & 0.089 & 0.068     & --      & $\infty$     \\
CaMgSi$_2$O$_6$:Eu            & 2.969 & 2.722 & 0.275 & 0.247 & 0.083 & -0.027    & 4.762   & 3.488   \\
SrAl$_2$O$_4$-Eu2             & 3.167 & 2.892 & 0.345 & 0.275 & 0.087 & -0.071    & 3.846   & 2.236   \\
Sr$_5$(PO$_4$)$_3$Cl-Eu       & 3.238 & 3.094 & 0.169 & 0.144 & 0.044 & -0.025    & 7.030  & 5.243  \\
CaF$_2$:Eu                    & 3.257 & 3.159 & 0.114 & 0.098 & 0.030 & -0.016    & 9.346   & 6.823   \\
SrI$_2$:Eu                    & 3.349 & 3.227 & 0.089 & 0.122 & 0.036 & 0.033     & --      & $\infty$     \\
Sr$_2$MgSi$_2$O$_7$:Eu        & 3.107 & 2.852 & 0.329 & 0.255 & 0.082 & -0.074    & 3.876   & 2.144   \\
SrAl$_2$O$_4$-Eu1             & 2.968 & 2.677 & 0.361 & 0.291 & 0.098 & -0.071    & 3.559   & 1.907   \\
BaSi$_2$O$_2$N$_2$-Eu         & 2.855 & 2.607 & 0.310 & 0.248 & 0.087 & -0.063    & 3.876   & 2.04   \\
Ba$_3$Si$_6$O$_{12}$N$_2$-Eu  & 2.940 & 2.864 & 0.100 & 0.076 & 0.019 & -0.025    & 9.901   & 6.102   \\
CaAlSiN$_3$:Eu, M-I           & 2.367 & 2.196 & 0.168 & 0.171 & 0.072 & 0.003     & 7.353   & 6.872   \\
CaAlSiN$_3$:Eu, M-II          & 2.389 & 2.195 & 0.164 & 0.173 & 0.074 & 0.003     & 7.353   & 6.872   \\
Sr[Mg$_3$SiN$_4$]:Eu          & 2.216 & 2.140 & 0.085 & 0.076 & 0.034 & -0.008    & 9.615   & 5.634   \\
CaS:Eu                        & 2.120 & 1.976 & 0.166 & 0.144 & 0.065 & -0.022     & 5.253  & 2.601   \\   
Sr[LiAl$_3$N$_4$]:Eu1         & 2.095 & 2.038 & 0.076 & 0.057 & 0.027 & -0.019    & 7.937   & 2.727   \\
Sr[LiAl$_3$N$_4$]:Eu2         & 2.160 & 2.049 & 0.060 & 0.111 & 0.051 & 0.052     & --      & $\infty$     \\                
Ca[LiAl$_3$N$_4$]:Eu          & 1.992 & 1.910 & 0.087 & 0.082 & 0.041 & -0.005    & 9.346   & 5.638   \\
\hline
\end{tabular}
\end{table*}

\subsection{Full Width at Half-Maximum and Temperature Shift}
\label{FWTH}

\begin{figure}[h]
\includegraphics[scale=0.25]{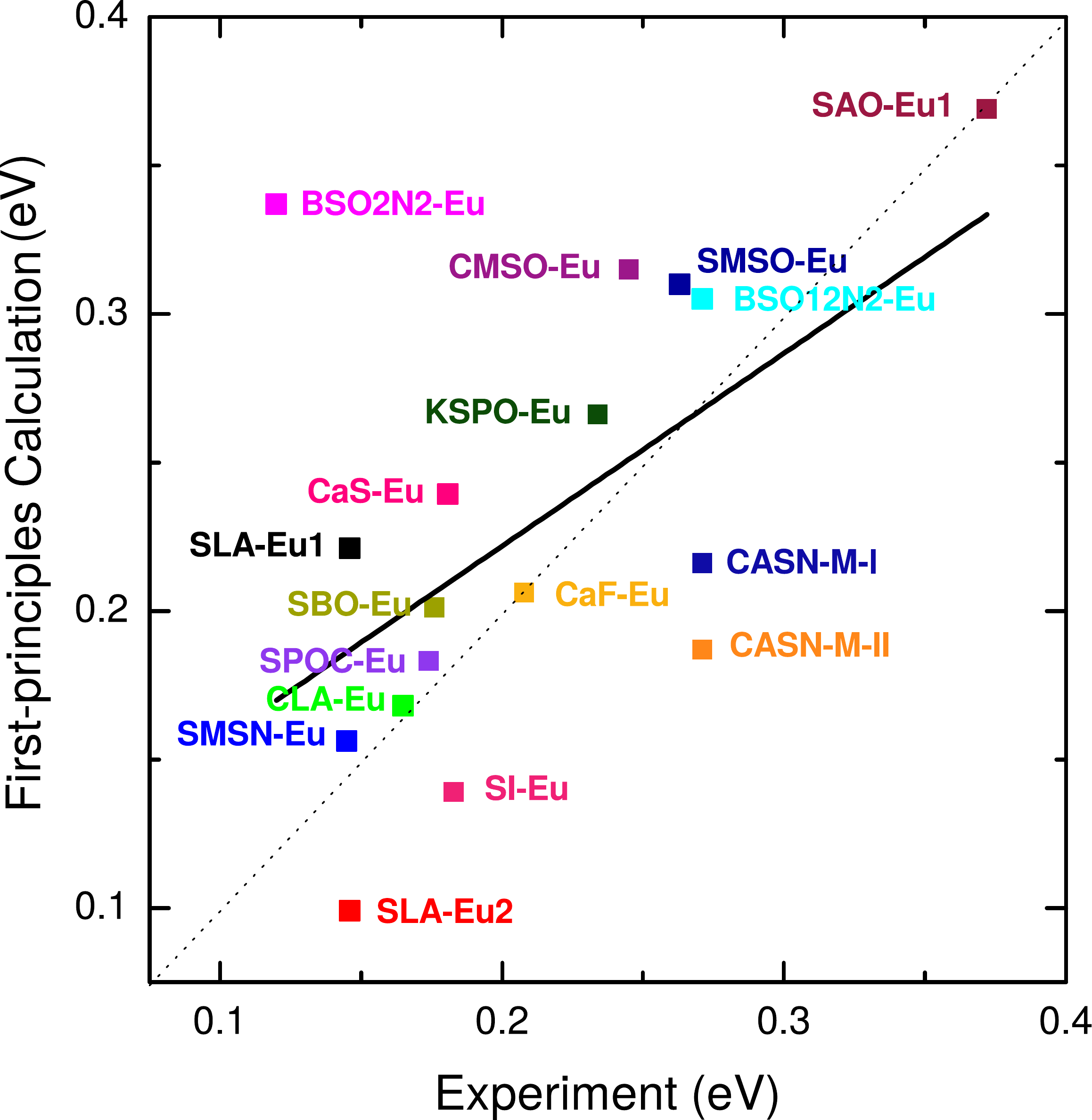}
\caption{Comparison of the FWHM between first-principles calculations and experiment at room temperature.}\label{figure_5}
\end{figure}

Following the methodology mentioned in Section~\ref{conf_coord}, we have computed the parameters of the luminescence spectrum shape line for the Eu$^{2+}$-doped materials.
The results are listed in Table~\ref{table-4}. 
Fig.~\ref{figure_5} shows the comparison between the experimental and theoretical results for the FWHM at room temperature.
The statistical analysis of the results is shown in Table~\ref{table-2}, the rightmost column. 
The average absolute error of the room temperature FWHM with respect to experimental data is around 0.05 eV for a range of experimental values of 0.1-0.4 eV. There is reasonable predictive power in this approach.

However, discrepancies for selected cases (in bold) are nearly twice as big.
The largest deviations are found for BaSi$_2$O$_2$N$_2$:Eu (theory overestimates the experimental 
FWHM by a factor of three) and CaAlSiN$_3$:Eu (theory underestimates the experimental FWHM by 30\%).
The theoretical underestimation of the FWHM might be due to different available phase structures or disorder. In particular, for CaAlSiN$_3$:Eu, we assume that the discrepancy is due to cations partial occupancy in the crystal structure,\cite{CASN1,CASN2} which leads to an inhomogeneous broadening of the emission peak. In the present work, the two ordered structures might fail to describe the complex environment  surrounding the Eu$^{2+}$ site, resulting in a smaller FWHM compared to experiment. 

\begin{figure}
\includegraphics[scale=0.20]{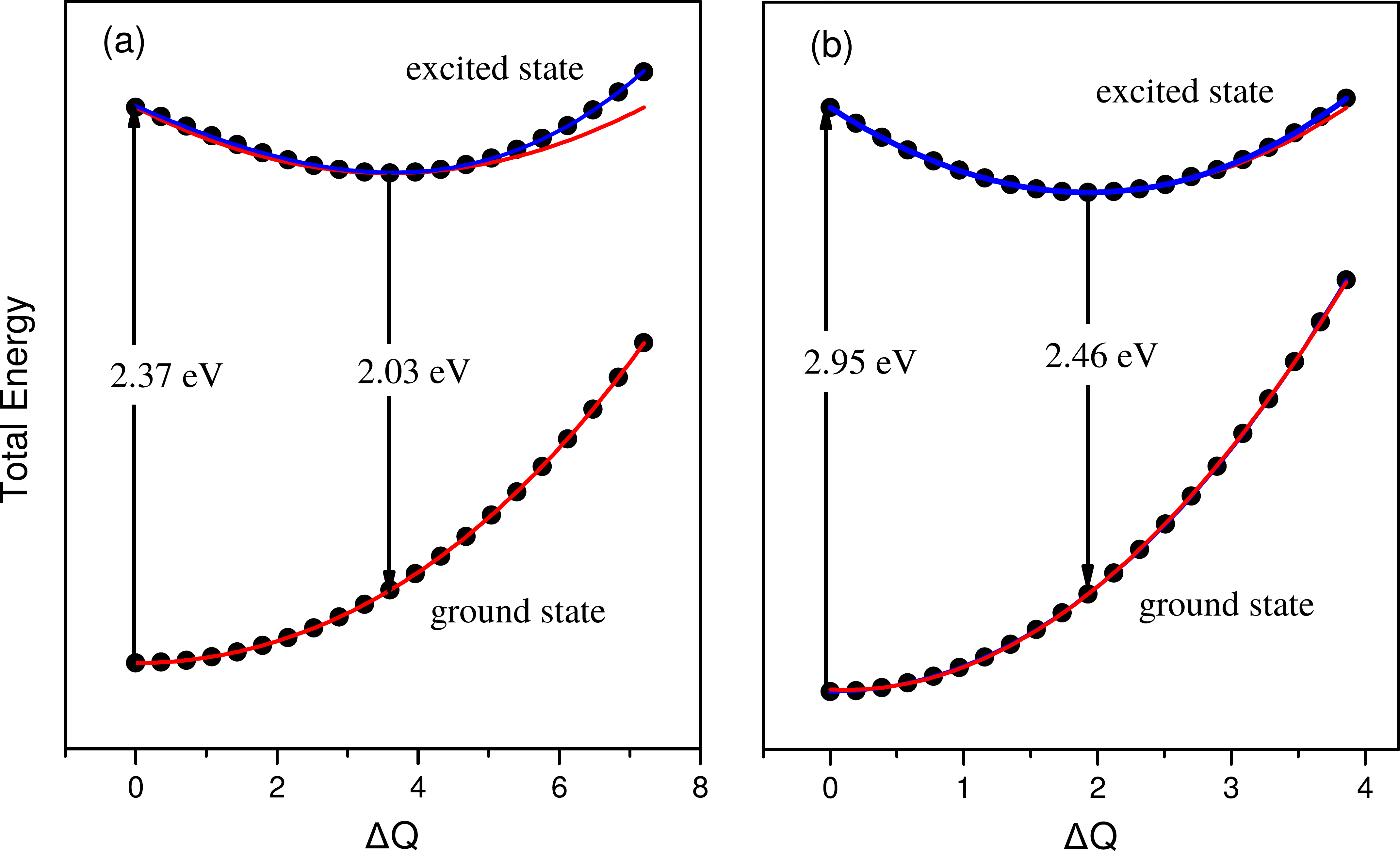}
\caption{Energy curves of the ground and excited states: (a) CaAlSiN$_3$:Eu with M-I;  (b) Ba$_3$Si$_6$O$_{12}$N$_2$:Eu. The red line is a fit using a second-order polynomial
with constrained energy at $Q_g$ and $Q_e$ and constrained location of  the minimum of the curve, while the blue line is a least-square fit using a third-order polynomial. Anharmonicities are weak for the ground state as well as for the Ba$_3$Si$_6$O$_{12}$N$_2$:Eu excited state, but more important
for CaAlSiN$_3$:Eu with M-I. However, even in this case, the effect develops mainly beyond $Q_e$, and modify the estimated FWHM by 6\%, as discussed in the text.}\label{figure_6}
\end{figure}

We have also tested the possibility that non-harmonic effects could modify significantly the 
theoretical predictions for such compounds, nevertheless relying on the 1D-CCD methodology. Fig.~\ref{figure_6} shows the potential energy curves of the ground and excited states in the CaAlSiN$_3$:Eu and Ba$_3$Si$_6$O$_{12}$N$_2$:Eu, as a function of the $Q$ coordinate, as described in Sec. IIC.  These energy curves have been fitted by a second-order polynomial curve
with constrained energy at $Q_g$ and $Q_e$ also constraining the location of  the minimum of the curve, while the blue line is a least-square fit using a third-order polynomial curve.  Table~\ref{table-5} lists the fitting parameters. As discussed in Sec. IIC, the FWHM from non-harmonic effect  can be deduced from the ratio between the second derivatives at the minimum of the third-order polynomial curve and of the second-order polynomial curve, to the power -1/4. For the CaAlSiN$_3$:Eu and  Ba$_3$Si$_6$O$_{12}$N$_2$:Eu, the ratio between FWHM is calculated to be 0.94 and 0.99, respectively, which is 
quite close to unity and indicate the relative smallness of non-harmonic effect for the two compounds. Therefore, the change of FWHM from non-harmonic effect should be much smaller than the disorder effect in CaAlSiN$_3$:Eu.  

The reasons for the large errors in Sr[LiAl$_3$N$_4$]:Eu and CaMgSi$_2$O$_6$:Eu are not clear at the moment. The description of the FWHM as well as thermal quenching behaviour in these four systems might require to go beyond the 1D-CCD, or to resort to a more advanced DFT approximation than the GGA, 
e.g. hybrid functionals.

At present, experimental FWHM data at low temperature (4~K) is not available for most of the systems. Therefore, the corresponding comparison between experiment and theory need further experimental contribution in the low temperature region.    

Beside the information on the FWHM at low temperature, the 1D-CCD yields the modification of the spectrum shape with temperature, see Eq.~\eqref{tem}.
Most of Eu$^{2+}$-doped materials show blue shift with higher temperature. 
So far, the effect of spin-orbit coupling for the excited state of Eu$^{2+}$ ion has not been considered. 

\begin{table*}[hbtp]
\centering
\renewcommand\arraystretch{1.5}
\caption{Analysis of luminescence line width of Eu$^{2+}$-doped phosphors. W(0~K) and W(298~K) stand for the calculated FWHM at 0~K and 298~K. 
The experimental data is at 298~K. $\Delta$E denotes the energy shift of emission peak at 298~K compared to the result at 0~K. 
Positive values denote blue shift and negative value means a redshift with temperature. The numbers in bold deviate substantially from experiment.}
\label{table-4}
\begin{tabular}{lcccccccccccc}
\hline
Compound & $\Delta Q$          & $\Delta \text{R}$        & M      & $\displaystyle \hbar \Omega_\text{g}$  & $\displaystyle \hbar \Omega_\text{e}$  & S$_{\text{abs}}$    & S$_{\text{em}}$     & W(0K) & W(298K) & W(Exp)   & $\Delta$E & Exp. Ref. \\
& [amu$^{1/2}$*\AA]         & [\AA]        & [amu]     & [meV]  & [meV]  &     &      & [eV] & [eV] & [eV]   & [eV]\\\hline
SrB$_4$O$_7$:Eu	&	1.048 	&	0.208 	&	25.28 	&	28.0 	&	28.8 	&	3.78 	&	3.69 	&	0.125 	&	0.201 	&	0.176 	&	0.004 	 & [\onlinecite{SBO}]\\
KSrPO$_4$:Eu	&	5.739 	&	1.001 	&	32.86 	&	8.8 	&	9.7 	&	38.18 	&	34.51 	&	0.115 	&	0.266 	&	0.234 	&	0.010 	 & [\onlinecite{KSPO}] \\
CaMgSi$_2$O$_6$:Eu	&	1.323 	&	0.295 	&	20.15 	&	36.3 	&	34.4 	&	7.20 	&	7.57 	&	0.241 	&	\textbf{0.315} 	&	0.245 	&	0.028  & [\onlinecite{CMSO}]	\\
SrAl$_2$O$_4$-Eu2	&	2.942 	&	0.530 	&	30.86 	&	18.3 	&	16.3 	&	16.86 	&	18.88 	&	0.198 	&	0.358 	&	- 	&	0.041  & [\onlinecite{SAO}]	\\
Sr$_5$(PO$_4$)$_3$Cl-Eu	&	2.737 	&	0.480 	&	31.56 	&	13.7 	&	12.7 	&	11.37 	&	12.30 	&	0.118 	&	0.240 	&	0.174 	&	0.019  & [\onlinecite{SPOC}]	\\
CaF$_2$:Eu	&	0.955 	&	0.219 	&	19.00 	&	32.3 	&	29.9 	&	3.27 	&	3.54 	&	0.149 	&	0.206 	&	0.208 	&	0.013 	 & [\onlinecite{CaF2}]\\
SrI$_2$:Eu	&	3.657 	&	0.329 	&	123.36 	&	7.5 	&	8.8 	&	13.99 	&	11.98 	&	0.057 	&	0.139 	&	0.183 	&	-0.002  & [\onlinecite{SrI2}]	\\
Sr$_2$MgSi$_2$O$_7$:Eu	&	2.215 	&	0.462 	&	22.98 	&	23.7 	&	20.8 	&	12.26 	&	13.89 	&	0.221 	&	0.310 	&	0.263 	&	0.040 	 & [\onlinecite{SMSO}]\\
SrAl$_2$O$_4$-Eu1	&	2.273 	&	0.450 	&	25.48 	&	23.7 	&	21.7 	&	13.41 	&	15.26 	&	0.233 	&	0.369 	&	0.372 	&	0.042  & [\onlinecite{SAO}]	\\
BaSi$_2$O$_2$N$_2$-Eu	&	3.447 	&	0.397 	&	75.50 	&	14.8 	&	13.2 	&	18.77 	&	21.00 	&	0.169 	&	\textbf{0.337} 	&	0.120 	&	0.042  & [\onlinecite{BSO2N2}]	\\
Ba$_3$Si$_6$O$_{12}$N$_2$-Eu 	&	1.930 	&	0.408 	&	22.43 	&	24.5 	&	22.8 	&	10.16 	&	10.88 	&	0.197 	&	0.305 	&	0.271 	&	0.029  & [\onlinecite{BSO12N2}]	\\
CaAlSiN$_3$, M-I	&	3.611 	&	0.663 	&	29.69 	&	10.4 	&	10.5 	&	16.49 	&	16.08 	&	0.097 	&	\textbf{0.216} 	&	0.278 	&	0.016  & [\onlinecite{CASN}]	\\
CaAlSiN$_3$, M-II	&	2.635 	&	0.475 	&	30.76 	&	13.1 	&	14.2 	&	11.84 	&	10.84 	&	0.097 	&	\textbf{0.187} 	&	0.278 	&	0.008  & [\onlinecite{CASN}]	\\
Sr{[}Mg$_3$SiN$_4${]}:Eu	&	1.772 	&	0.258 	&	38.59 	&	15.0 	&	14.2 	&	5.35 	&	5.62 	&	0.081 	&	0.156 	&	0.145 	&	0.012  & [\onlinecite{SMSN}]	\\
CaS:Eu	&	2.166 	&	0.381 	&	32.29 	&	17.2 	&	16.0 	&	9.01 	&	9.65 	&	0.130 	&	0.237 	&	0.181 	&	0.024  & [\onlinecite{CaS}]	\\
Sr[LiAl$_3$N$_4$]:Eu1	&	0.756 	&	0.172 	&	19.28 	&	33.4 	&	29.0 	&	1.71 	&	2.63 	&	0.158 	&	\textbf{0.221} 	&	0.146 	&	0.027  & [\onlinecite{SLA}]	\\
Sr[LiAl$_3$N$_4$]:Eu2	&	1.222 	&	0.143 	&	73.01 	&	18.3 	&	25.0 	&	4.46 	&	3.27 	&	0.067 	&	0.099 	&	0.146 	&	-0.009  & [\onlinecite{SLA}]	\\
Ca[LiAl$_3$N$_4$]:Eu	&	1.396 	&	0.259 	&	28.97 	&	19.3 	&	18.6 	&	4.35 	&	4.52 	&	0.099 	&	0.168 	&	0.165 	&	0.013  & [\onlinecite{CLA}]	\\
\hline
\end{tabular}
\end{table*}

\begin{table*}
\centering
\renewcommand\arraystretch{1.5}
\caption{Parameters of the fitted polynomials in Fig.~\ref{figure_s}: $E = C_2Q ^2 + C_1 Q + C_0$ for the second-order polynomials, and $E = C_3 Q^3 + C_2 Q^2 + C_1 Q + C_0$ for third-order polynomials.}
\label{table-5}
\begin{tabular}{l|c|c|c|c|c|c|c|c}
\hline\hline

Compound & State  & Polynomial & $C_3$ & $C_2$ & $C_1$   & $Q_{\rm min}$ & $E(Q_{\rm min})$[Ha] & $\frac{\partial^2E}{\partial Q^2}|_{Q_{\rm min}}$\\ \hline
\multirow{ 4}{*}{CaAlSiN$_3$:Eu } & Ground state  & Second-order &- & 4.51E-4 & 0  & 0 & -647.428 & 9.02E-4 \\
& Ground state  & Third-order & 1.44E-5 & 3.83E-4 & 6.19E-5  & 0 & -647.428 & 7.66E-4 \\ 
& Excited state  & Second-order & - & 4.07E-4 & -2.93E-3  & 3.611 & -647.347 & 8.15E-4 \\ 
& Excited State  & Third-order    & 3.28E-5 & 1.65E-4 & -2.51E-3 & 3.611 & -647.346 & 10.40E-4 \\ \hline
\multirow{ 4}{*}{Ba$_3$Si$_6$O$_{12}$N$_2$:Eu} & Ground State  & Second-order & - & 2.63E-3 & 0  & 0 & -1020.339 & 5.26E-3 \\
& Ground State  & Third-order &7.82E-5 & 2.46E-3 & 3.01E-5 & 0 & -1020.339 &  4.92E-3 \\ 
& Excited State  & Second-order & - & 2.29E-3 & -8.81E-3 & 1.930 & -1020.289 & 4.58E-3 \\ 
& Excited State  & Third-order & -1.17E-13 & 2.41E-3 & 9.14E-3 & 1.930 & -1020.280 & 4.82E-3 \\  \hline\hline
\end{tabular}
\end{table*}


\subsection{Fitting the Dorenbos' semi-empirical model}
\label{Fitting}

The accuracy of the $\Delta$SCF method has been assessed by direct comparison between the calculated transition energies and Stokes shifts and the corresponding experimental data
for the fifteen representative Eu$^{2+}$-doped phosphors. 
In addition, the relaxed structures for the ground and excited states have been obtained. In this section, such structural information is used to fit the semi-empirical model parameters
for the Eu$^{2+}$ ion proposed in the Section~\ref{Dorenbos}, and assess its predictive strength.

\begin{table}[h]
\centering \renewcommand\arraystretch{1.5}
\caption{Absorption/emission energy (eV) and Stokes shift (cm$^{-1}$), from the semi-empirical approach as well as from experiment. 
The numbers in bold deviate substantially from experiment, above 0.3 eV and 1000 cm$^{-1}$ for the transition energies and Stoke shift, respectively.} \label{table-6} \begin{tabular}{l|ccc|cccc} 
\hline\hline
 Compound &\multicolumn{3}{c|}{Semi-empirical}&\multicolumn{4}{c}{Experiment} \\
 & Abs   &Em   &$\Delta$S &Abs   &Em   &$\Delta$S & Ref. \\ \hline
 SrB$_4$O$_7$:Eu & 3.375 & 3.329 & \textbf{374} & 3.54 & 3.35 & 1502  & [\onlinecite{SBO}]\\
 KSrPO$_4$:Eu & 3.650 & 2.731 & \textbf{7408} & 3.32 & 2.88 & 3500  & [\onlinecite{KSPO}]\\  
 CaMgSi$_2$O$_6$:Eu & 2.902 & 2.712 & \textbf{1534} & 3.16 & 2.72 & 3188  & [\onlinecite{CMSO}]\\ 
 SrAl$_2$O$_{4}$:Eu2 & 3.135 & 2.881 & 2056 & 3.11 & 2.79 & 2581  & [\onlinecite{SAO}]\\  
 CaF$_2$:Eu & 3.270 & 3.190 & 645 & 3.06 & 2.92 & 1047  & [\onlinecite{CaF2}]\\ 
 Sr$_2$MgSi$_2$O$_7$:Eu & 3.225 & 3.070 & 1251 & 2.94 & 2.70 & 1936  & [\onlinecite{SMSO}]\\ 
 SrAl$_2$O$_{4}$:Eu1 & 3.143 & \textbf{2.963} & \textbf{1454} & 2.88 & 2.38 & 4033  & [\onlinecite{SAO}]\\
 BaSi$_2$O$_{2}$N$_2$:Eu & 2.924 & 2.630 & 2369 & 2.71 & 2.52 & 1532  & [\onlinecite{BSO2N2}]\\
 Ba$_3$Si$_6$O$_{12}$N$_2$:Eu & \textbf{3.327} & \textbf{3.155} & \textbf{1390} & 2.69 & 2.32 & 2790  & [\onlinecite{BSO12N2}]\\ 
 Sr[Mg$_3$SiN$_4$]:Eu  & 2.180 & 2.016 & 1321 & 2.26 & 2.02 & 1935  & [\onlinecite{SMSN}]\\
 Sr[LiAl$_3$N$_4$]:Eu1 & 1.962 & 1.809 & 1237 & 2.03 & 1.91 & 956  & [\onlinecite{SLA}] \\
 Sr[LiAl$_3$N$_4$]:Eu2 & 1.971 & 1.819 & 1230 & 2.03 & 1.91 & 956  & [\onlinecite{SLA}] \\
 Ca[LiAl$_3$N$_4$]:Eu  & 1.838 & 1.653 & 1490 & 1.96 & 1.86 & 800  & [\onlinecite{CLA}]\\
\hline\hline 
\end{tabular} \end{table}

\begin{figure*}[hbtp]
\includegraphics[scale=0.25]{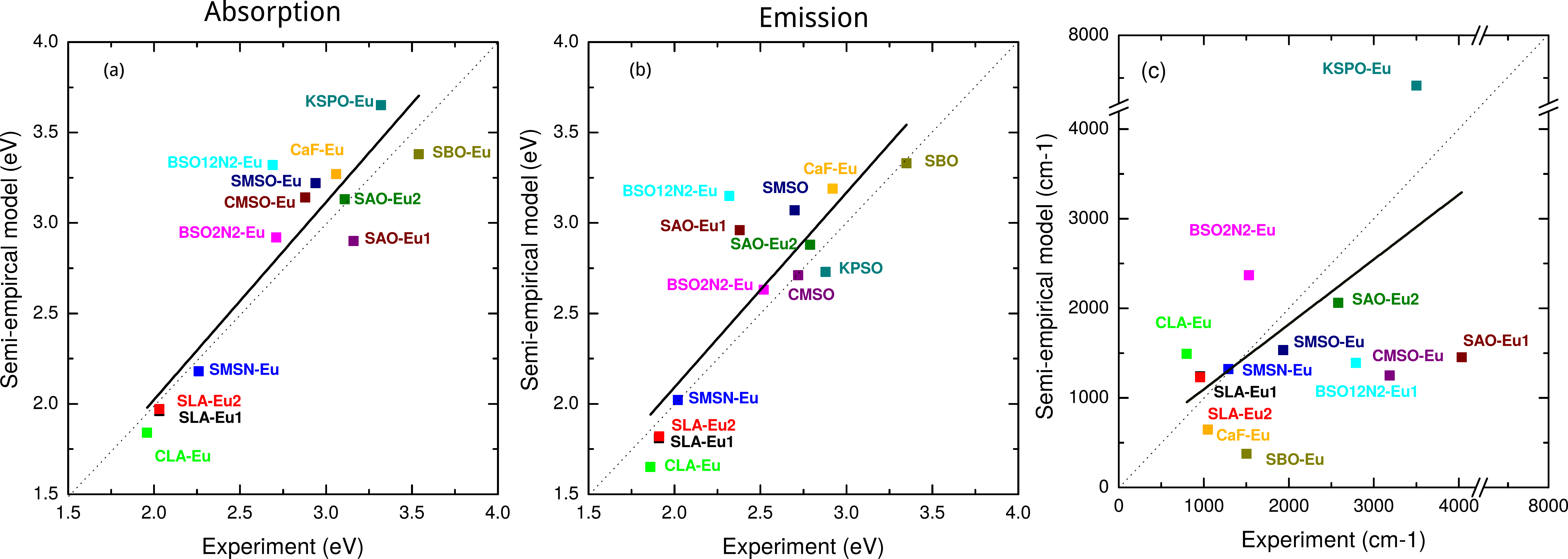}
\caption{Comparison between experimental results and semi-empirical calculations: (a) absorption energy, (b) emission energy and (c) Stokes shift. }\label{figure_7}
\end{figure*}

For the quantitative determination of the semi-empirical model, two parameters are needed, the spectroscopic polarization $\alpha_{sp}$ and the crystal-field splitting $\beta$. 
Among the fifteen compounds studied in Sec.~\ref{validation}, the spectroscopic polarization $\alpha_{sp}$ is only available for twelve of them (oxides, nitrides and one fluoride) 
and the $\beta$ parameter for tetrahedral coordination environment is missing for CaAlSiN$_{3}$:Eu. 
Therefore, eleven compounds have been selected here for the analysis of the semi-empirical model. 
The detailed information for the determination of the redshift is given in the Appendix and the calculated transition energies and Stokes shifts from the semi-empirical model are shown in Table~\ref{table-6}. 
The calculated transition energies and Stokes shifts from the semi-empirical model matches experiment within 0.3 eV in most cases, 
while for the cases of Ba$_3$Si$_6$O$_{12}$N$_2$:Eu, SrAl$_2$O$_{4}$:Eu1 and KSrPO$_4$:Eu, the error is larger.
For the Stokes shift, the semi-empirical method gives a larger error for SrB$_4$O$_7$:Eu and KSrPO$_4$:Eu. 
Fig. \ref{figure_7} shows the direct comparison between the semi-empirical model and experiment. 
The corresponding statistical analyses, examining a linear relationship between theory and experiment, have been performed. 
Detailed information is shown in Table~\ref{table-7}.  
Reasonable results for most cases were obtained and indeed showed the predicting capability of the proposed semi-empirical method, 
while the above-mentioned limitation indicated that some additional work is needed on this model. 

\begin{table}
\centering
\renewcommand\arraystretch{1.5}
\caption{Statistical analysis of transition energies (eV) and Stokes shift $\Delta$S (cm$^{-1}$) from the semi-empirical model. 
ME, MAE, MRE and MARE stand for the mean error, mean absolute error, mean relative error, and mean absolute relative error, respectively. 
The slope, intercept and coefficient of determination (R-Square) are determined by the fitting lines shown 
in Fig.~\ref{figure_7} for the transition energy. }
\label{table-7}
\begin{tabular}{l|ccc}
\hline\hline
 &\multicolumn{3}{c}{Semi-empirical model} \\
 & Absorption & Emission& Stokes shift   \\ \hline
ME   & 0.091 eV & -0.011 eV & -181  \\
MAE  & 0.207 eV & 0.137 eV & 1107  \\
MRE (\%) & 2.95 & -0.43 & 1.27 \\
MARE (\%)& 7.39 & 5.18 & 49.4 \\
Slope   & 1.091 & 1.075 & 0.725  \\
Intercept& -0.160 & -0.059 & 372 \\ 
R-Square (\%)& 83.0 & 70.5 & 13.1  \\ \hline\hline
\end{tabular}
\end{table}

\section{Conclusion}
\label{conclusion}
In this work, the luminescence characteristics of fifteen representative Eu$^{2+}$-doped materials have been systematically studied from first principles. 
The $\Delta$SCF methodology, with CDFT (GGA-PBE+U) has been used to describe the ground- and excited- states of the Eu$^{2+}$ ion, from which transition energies and Stokes shift have been deduced. 
For all fifteen materials, the calculated transition energies match experiment within 0.3~eV, the $4f$ levels are located in the gap in the ground-state electronic structure,  the upper (unoccupied) $4f$ and lower (occupied) $5d$ are also located in the gap in the excited (CDFT) state. This is however not the case for CaO:Eu, for which the method fails, as the $5d$ state does not enter the band gap when such methodology is followed.
 
The slope of the least-square fitting line adjusted to the experimental versus theoretical absorption and emission lines is close to unity, while the
intercept is reasonably close to zero.
The information on total energies and structure geometry were then used to provide an estimate of the energy barrier for the thermal quenching 
via $4f-5d$ crossover (E$_{\text{fd}}$) and FWHM of the emission band following the 1D-CCD. 
For the E$_{fd}$, the calculated value for all the Eu$^{2+}$-doped phosphors is above 1.5~eV, 
indicating that the auto-ionization is likely the dominant mechanism for the thermal quenching behaviour. 
For the FWHM, the calculated values at room temperature match experiment at room temperature with an average absolute error of around 0.05 eV, 
for a range of experimental values between 0.120~eV and 0.372~eV,
despite the use of the crude 1D-CCD model for the analysis of electron-vibrational coupling. 
Finally, parameters from first-principles geometries (e.g. average nearest-neighbours) have been extracted, and used in a semi-empirical model. 
The obtained transition energies from this semi-empirical model were compared to the experimental data, 
giving an error above 0.5~eV for two of the materials. The predicting power of the semi-empirical model is found to be more limited in its accuracy and scope than the first-principles method.
Its interest lies in the identification of the origin of the variation of absorption and emission energies
and the physical interpretation of different geometrical quantities or polarisation of the ions.

\begin{acknowledgments}
We thank J.-M. Beuken for support in the use of computational resources.
This work, done in the framework of ETSF (project number 551),
has been supported by the Fonds de la Recherche Scientifique (FRS-FNRS Belgium) through the PdR Grant No. T.0238.13 - AIXPHO.
Computational resources have been provided by the supercomputing facilities
of the Universit\'e catholique de Louvain (CISM/UCL)
and the Consortium des Equipements de Calcul Intensif en
F\'ed\'eration Wallonie Bruxelles (CECI) funded by the FRS-FNRS under Grant No. 2.5020.11.
\end{acknowledgments}


\appendix*
\renewcommand\thefigure{A\arabic{figure}}
\setcounter{figure}{0}
\renewcommand\thetable{A\arabic{table}}
\setcounter{table}{0}   
\section{}

The calculation parameters for the fifteen Eu$^{2+}$-doped materials are shown in this Appendix. 
These include the structural description for several compounds as well as more detailed results for the SLA:Eu phosphor.

\subsection{Supercell calculation} 

\begin{figure}
\includegraphics[scale=0.7]{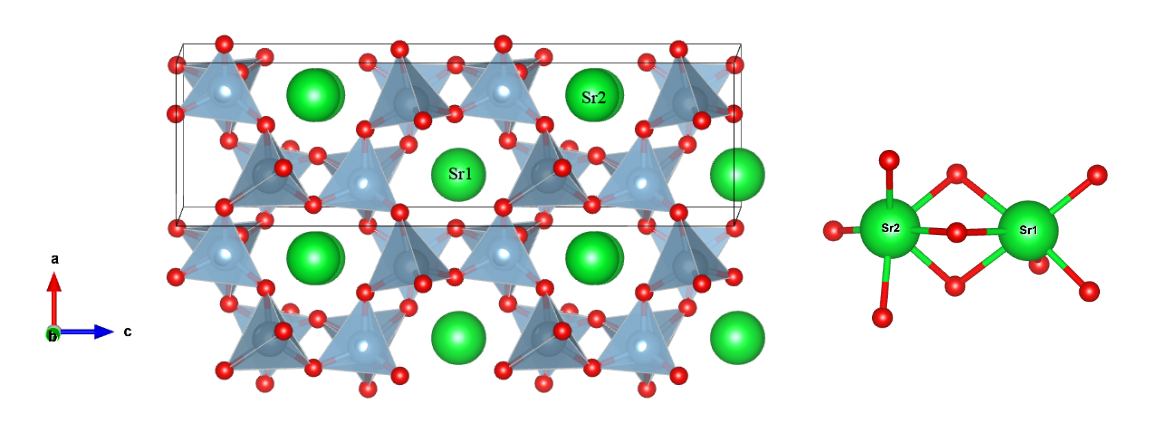}
\caption{Crystal structure of SrAl$_2$O$_4$. The grey,  green and red spheres stand for Al,Sr and O atoms, respectively.}\label{figure_a1}
\end{figure}

\begin{figure}
\includegraphics[scale=0.35]{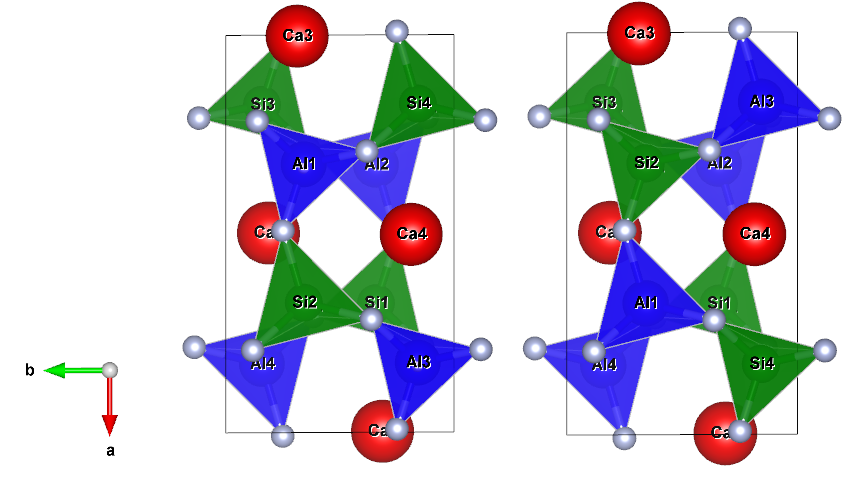}
\caption{Crystal structure of CaAlSiN$_3$ for the first (M-I, left) and second (M-II, right) model (see text for more information on the two models). The green, blue, red and grey spheres stand for the Si, Al, Ca and
N atoms, respectively.}\label{figure_a2}
\end{figure}

In this work, the first-principles calculations have been conducted using the supercell method. 
This method can take the deformation of the crystal structure in the ground- and excited- states into account, 
which is required for the calculation of the luminescence spectrum line shape and the FWHM. 
For each of the fifteen compounds, a detailed convergence study on the supercell size and \textbf{k}-point sampling has been conducted, and the parameters needed to obtain convergence within 0.1~eV for the transition energy, are shown in Table~\ref{table_A1}. 
There are four materials with multi-site possibilities for the Eu$^{2+}$ ion doping. 
For Ba$_3$Si$_6$O$_{12}$N$_2$:Eu, the assignment of the Eu sites corresponds to the one of our previous work.\cite{samuel2016} 
For Sr[LiAl$_3$N$_4$]:Eu, SrAl$_2$O$_4$:Eu and Sr$_5$(PO$_4$)$_3$Cl, we follow the notations in the literature.\cite{SLA,SAO,SPOC}. Figure~\ref{figure_a1} shows the crystal structure of SrAl$_2$O$_4$ and the local coordination environment of two nonequivalent Sr$^{2+}$ sites.
For the CaAlSiN$_3$ compound, experimental results have shown that Al$^{3+}$ and Si$^{4+}$ cation ions are disordered in the crystal structure. 
Here, we use two symmetrical crystal models (M-I and M-II) to mimic the disordered structural geometry in Figure~\ref{figure_a2}, following the previous work of Mikami et al.\cite{CASN1} 

\begin{table}
\centering
\renewcommand\arraystretch{1.5}
\caption{The Brillouin Zone wavevector sampling, supercell size, and Eu-doping concentration of the fifteen Eu doped materials. 
The concentration refers to the total number of potential substitutional sites.}
\label{table_A1}
\begin{tabular}{cccc}
\hline\hline
Compound & \textbf{k}-point grid & Supercell size & Eu concentration
\tabularnewline
\hline
 SrB$_4$O$_7$:Eu & 2$\times$2$\times$2 & 96 & 12.5\%\\
 KSrPO$_4$:Eu & 2$\times$3$\times$2 & 56 & 12.5\%\\
 CaMgSi$_2$O$_6$:Eu & 2$\times$2$\times$2 & 80 & 12.5\%\\
 SrAl$_2$O$_4$:Eu2&  3$\times$3$\times$3 & 56 & 12.5\%\\
 Sr$_5$(PO$_4$)$_3$Cl:Eu&  3$\times$3$\times$2 & 84 & 5.00\%\\ 
 CaF$_2$:Eu & 2$\times$2$\times$2 & 48 & 6.25\%\\
 SrI$_2$:Eu & 4$\times$4$\times$2 & 48 & 6.25\%\\
 Sr$_2$MgSi$_2$O$_7$:Eu & 2$\times$2$\times$2 & 48 & 12.5\%\\
 SrAl$_2$O$_4$:Eu1&  3$\times$3$\times$3 & 56 & 12.5\%\\
 BaSi$_2$O$_2$N$_2$:Eu&  4$\times$6$\times$2 & 56 & 12.5\%\\     
 Ba$_3$Si$_6$O$_{12}$N$_2$:Eu &  2$\times$2$\times$2 & 69 & 11.1\%\\
 CaAlSiN$_3$:Eu,M-I & 2$\times$2$\times$4 & 48 & 12.5\%\\
 CaAlSiN$_3$:Eu,M-II & 2$\times$2$\times$4 & 48 & 12.5\%\\
 Sr[Mg$_3$SiN$_4$]:Eu & 3$\times$3$\times$3 & 72 & 12.5\%\\
 CaS:Eu& 2$\times$2$\times$2 & 64 & 3.13\%\\
 Sr[LiAl$_3$N$_4$]:Eu1& 4$\times$6$\times$4 & 72 & 12.5\%\\
 Sr[LiAl$_3$N$_4$]:Eu2& 4$\times$6$\times$4 & 72 & 12.5\%\\
 Ca[LiAl$_3$N$_4$]:Eu & 3$\times$3$\times$3 & 72 & 12.5\%\\
\hline\hline
\end{tabular}
\end{table}

\subsection{Calculation details of Sr[LiAl$_3$N$_4$]:Eu}

\begin{figure}[h]
\includegraphics[scale=0.3]{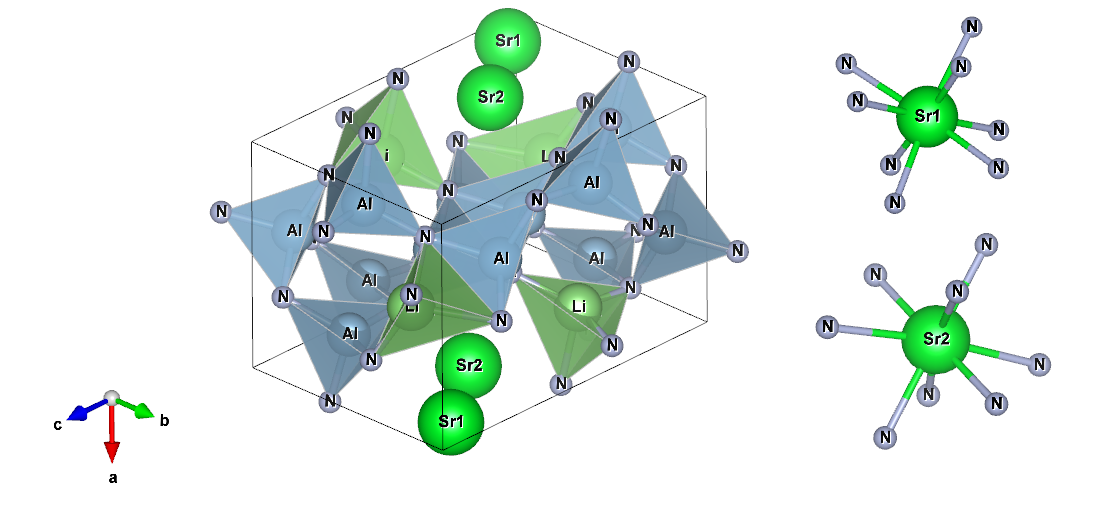}
\caption{Crystal structure of Sr[LiAl$_3$N$_4$]. The grey, light-blue, light-green and green spheres stand for N, Al, Li and Sr atoms, respectively.}\label{figure_a3}
\end{figure}

In this section, we show the calculation results for Sr[LiAl$_3$N$_4$]:Eu (SLA:Eu) as an example of the detailed outcome from the $\Delta$SCF method.  
The crystal structure of SLA is shown in Figure~\ref{figure_a3} which is a highly condensed, rigid framework of ordered edge- and corner-sharing AlN$_4$ and LiN$_4$ tetrahedra, 
with channels of four rings along [011]. These channels accommodate Sr$^{2+}$ ions to keep the neutral charge balance. 
There are two crystallographic Sr sites in this compound, each coordinated by eight N atoms in a highly symmetric cuboid-like environment. 
When doped, it can be expected that Eu$^{2+}$ ion will substitute the two Sr$^{2+}$ sites.
In the later, Eu1 and Eu2 stand for the Eu$^{2+}$ ion on the Sr1 and Sr2 site, respectively.

\begin{figure}[h]
\includegraphics[scale=0.5]{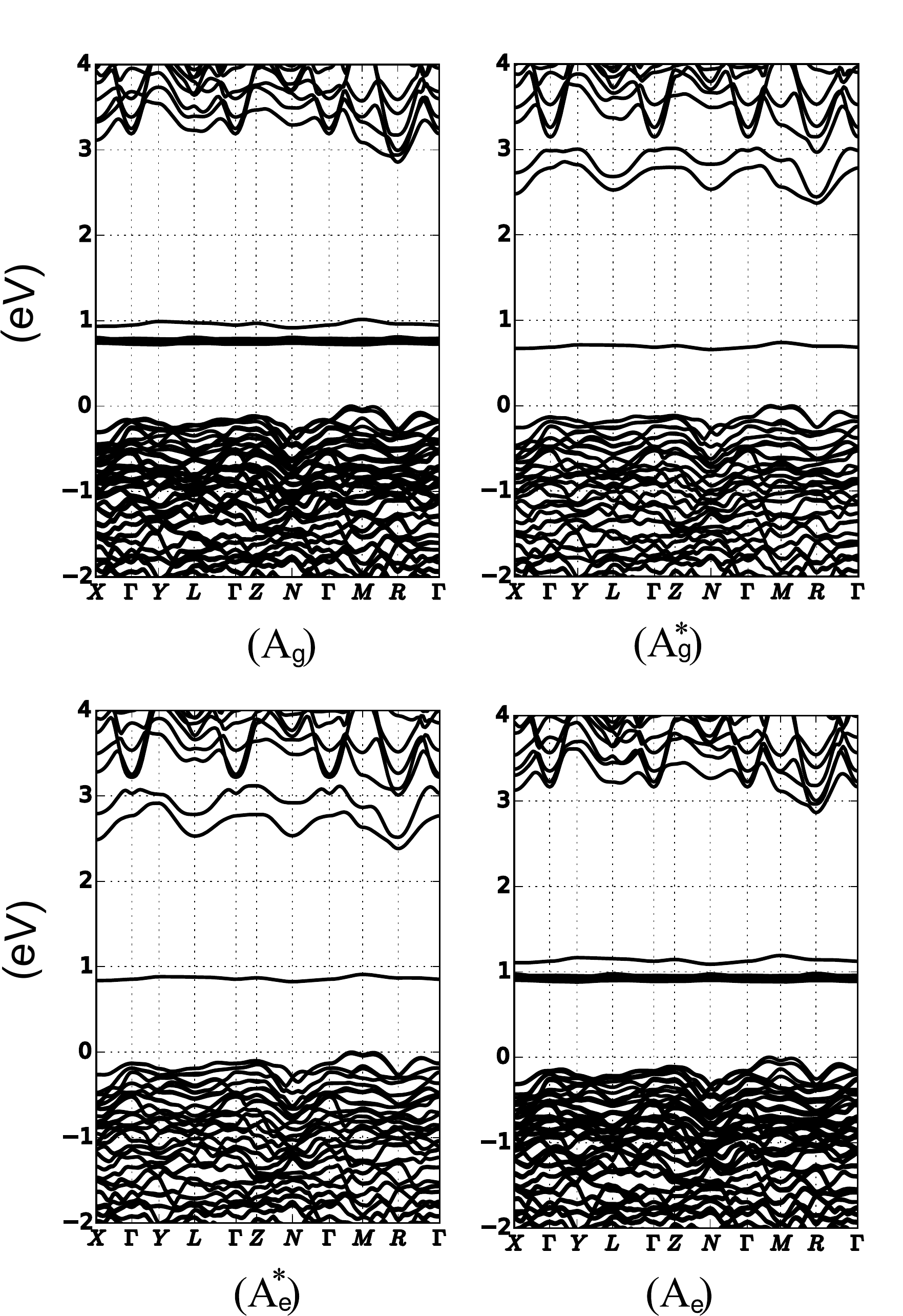}
\caption{Electronic band structures of SrLiAl$_3$N$_4$:Eu1. The meaning of A$_g$, A$_g^*$, A$_e^*$ and A$_e$ corresponds to the notation shown in the Figure~\ref{figure_2}. }\label{figure_a4}
\end{figure}

\begin{figure}[h]
\includegraphics[scale=0.53]{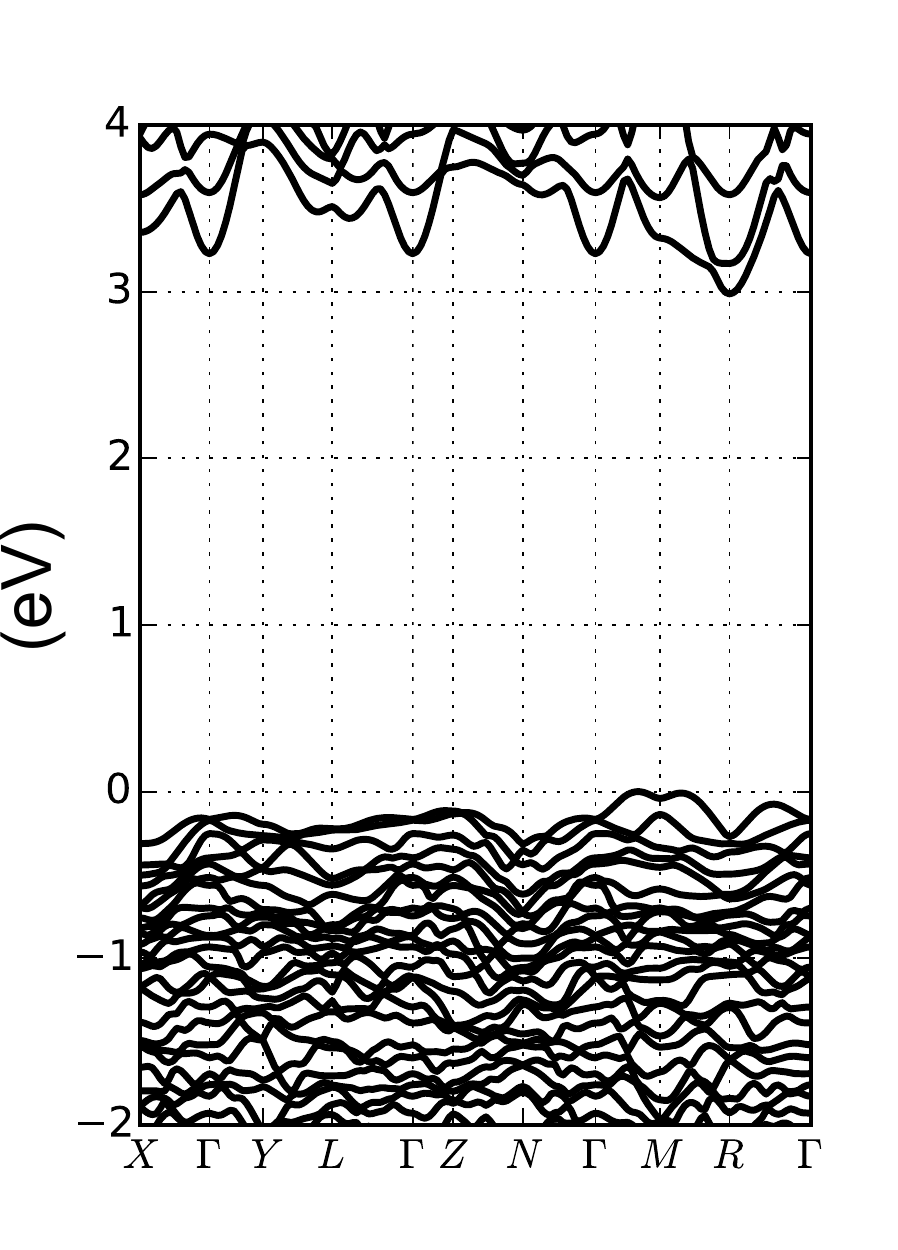}
\caption{Electronic band structures of the host SrLiAl$_3$N$_4$.}\label{figure_a5}
\end{figure}

As an example of what typically is observed for the fifteen Eu-doped materials, the electronic band structures of SLA:Eu1 are presented in Figure \ref{figure_a4} using the $\Delta$SCF method. 
The results of SLA:Eu2 are quite similar to those of SLA:Eu1.
In the ground state, there are seven flat bands occurring above the valence band maximum (VBM), not present in the undoped bulk, as shown in Figure~\ref{figure_a5}.
The shape of the CBM also changes with respect to the host calculation. 
In particular, its orbital content has changed from the Sr$_{4d}$ to the Eu$_{5d}$ state. 
Still, in the ground-state band structure, we can identify that some flat band constituted mostly by
Eu$_{5d}$ states just above the CBM.

\begin{figure}
\includegraphics[scale=0.04]{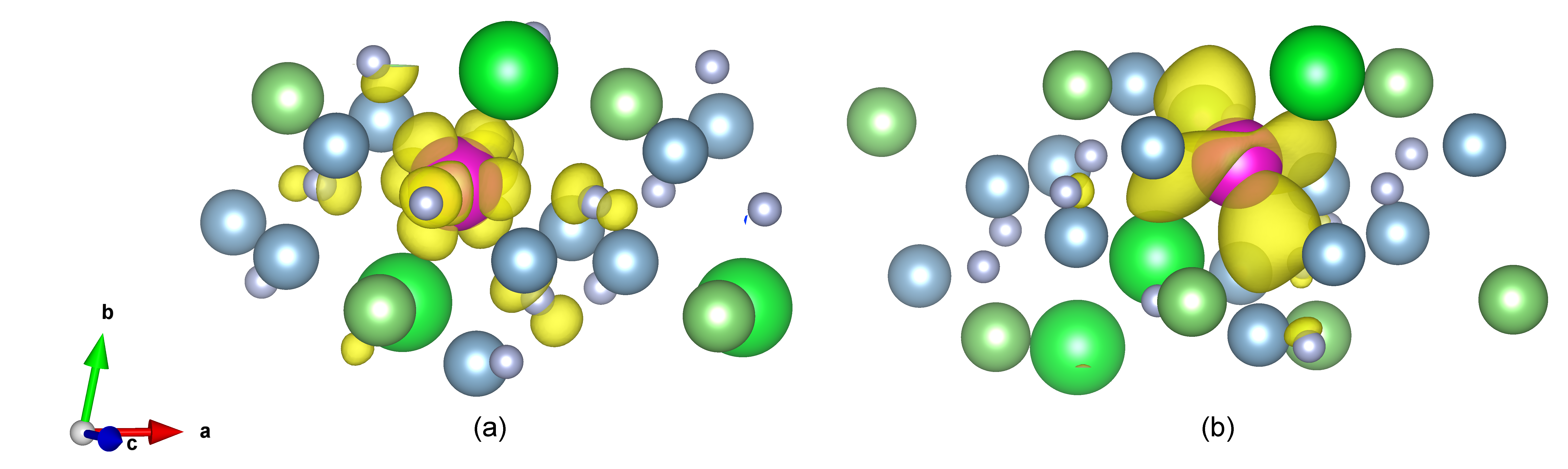}
\caption{Charge density of (a) LUMO and (b) HOMO of A$_e^*$ case of SrLiAl$_3$N$_4$:Eu1 at the $\Gamma$ point. The grey, light-blue, light-green, green and purple spheres stand for N, Al, Li, Sr and Eu atoms, respectively.}\label{figure_a6}
\end{figure}

In Fig.~\ref{figure_a6}, we present the LUMO and HUMO charge density of SLA:Eu1 at \textbf{k}=$\Gamma$ in the excited state A$_e^*$. 
There is only one unoccupied Eu$_{4f}$ state inside the band gap in the excited state, while seven occupied Eu$_{4f}$ states exist in the ground state.
From the analysis of the electronic band structure and charge density, we can see that the remaining six occupied Eu$_{4f}$ states are down-shifted into the valence band
when one Eu$_{4f}$ electron is promoted to the Eu$_{5d}$ state.
This is an artifact of the DFT+U formalism, to which little physical significance can be attributed. 
We also notice that the Eu$_{5d}$ states possess a low energy (large Eu$_{5d}$ to CBM gap) in the excited state due to the electron-hole interaction present in the $\Delta$SCF method.

Results for the two Sr[LiAl$_3$N$_4$]:Eu cases are presented in Table~\ref{table-A2}. We first note that 
both sites have similar total energies in the ground state A$_0$. Thus, the Eu$^{2+}$ ion equally substitutes the two Sr$^{2+}$ crystal sites.
The $\Delta$SCF method gives transition energies within 0.1~eV of the experimental data and within 30\% for the Stokes shift. 
The similar optical properties of the two inequivalent Eu$^{2+}$ substitution gives a narrow emission band. 

\begin{table}[hbtp]
\centering \renewcommand\arraystretch{1.5}
\caption{Transition energies and Stokes shift of Sr[LiAl$_3$N$_4$]:Eu. Experiment is from Ref~[\onlinecite{SLA}].}
\label{table-A2} \begin{tabular}{ccc} \hline
Case &  SLA:Eu1,12.5\% & SLA:Eu2,12.5\% \\ \hline
A$_g$ &  -20363.97 eV & -20363.96 eV \\
A$_g^{*}$&  -20361.87 eV & -20361.80 eV\\
A$_e^{*}$& -20361.93 eV& -20361.91 eV\\
A$_e$&  -20363.89 eV & -20363.90 eV \\\hline
$\Delta$E$_{abs}$(A$_g^{*}$-A$_g$)&  2.10 eV & 2.16 eV \\
$\Delta$E$_{abs}$(Exp.)&  2.03 eV & 2.03 eV  \\ 
$\Delta$E$_{em}$(A$_e^{*}$-A$_e$)&  1.96 eV & 1.99 eV \\
$\Delta$E$_{em}$(Exp.)& 1.91 eV & 1.91 eV \\ \hline
$\Delta$S(Cal.)& 1129 cm$^{-1}$ & 1371 cm$^{-1}$ \\
$\Delta$S(Exp.)& 956 cm$^{-1}$& 956 cm$^{-1}$  \\ \hline
\end{tabular} \end{table}

\bibliography{Eu_m}

\end{document}